\documentstyle[aps,amsfonts,epsf,floats]{revtex} \epsfverbosetrue
\let\isom = \cong
\let\implies = \Longrightarrow
\def\normsq#1{{{\left | #1 \right|}^2}}
\def\Nmap{{\normsq \cdot}}
\def\alg{{\frak A}}
\def\R{{\Bbb R}}
\def\C{{\Bbb C}}
\def\H{{\Bbb H}}
\def\O{{\Bbb O}}
\def\F{{\Bbb F}}
\def\K{{\Bbb K}}
\let\bra = \langle
\let\ket = \rangle
\def\bilin#1,#2{{\bra #1, #2 \ket}}
\def\blmap{{\bilin \cdot,\cdot}}
\def\Bar#1{{#1^*}}
\def\Star{{\Bar \cdot}}
\def\assoc#1,#2,#3{{[#1,#2,#3]}}
\def\ch#1{{\chi(#1)}}
\def\Zc{{\cal Z}}
\def\comm#1,#2{{[#1,#2]}}
\def\Comm#1,#2{{\big[#1,#2\big]}}
\def\iu{i}
\def\Z2{{\Bbb Z}_2}
\def\plane{\Z2 P^2}
\def\Psym{P}
\def\perm{{\Sigma}}
\def\oset{{\cal O}}
\def\sgr{H}
\def\tra{T}
\def\tsym{t}
\def\topp{\zeta}
\def\mapright#1{\;\smash{\mathop{\to}\limits^{#1}}\;}
\def\mapsrightto#1{\;\smash{\mathop{\mapsto}\limits^{#1}}\;}
\def\verylongrightarrow{\relbar\joinrel\longrightarrow}
\def\maplongright#1{\;\smash{\mathop{\longrightarrow}\limits^{#1}}\;}
\def\mapverylongright#1{\;\smash{\mathop{\verylongrightarrow}\limits^{#1}}\;}
\newsymbol\circlearrowleft 1309
\def\order#1{{|#1|}}
\def\orbit#1#2{{\text{Orb}_{#1}(#2)}}
\def\stab#1#2{{\text{Stab}_{#1}(#2)}}
\def\Tensor{{\cal T}}
\def\dmV{n}
\def\dm2V{m}
\def\ob{e}
\def\vol{\eta}
\def\opp{{\text{opp}}}
\def\quot#1/#2{\leavevmode\kern.1em\raise.5ex\hbox{$\textstyle #1$}
  \kern-.1em/\kern-.1em\lower.5ex\hbox{$\textstyle #2$}}
\def\vm{\mathbin{\lower.25ex\hbox{$\scriptscriptstyle\vee$}}}
\def\cpr_#1{\mathbin{%
  \mathchoice
    {\displaystyle{\mathop{\circ}_{\raise.7ex\hbox{$\scriptstyle#1$}}}}
    {\displaystyle{\mathop{\textstyle\circ}_%
       {\textstyle\raise.7ex\hbox{$\scriptstyle#1$}}}}
    {\displaystyle{\mathop{\scriptstyle\circ}_%
       {\scriptstyle\raise.7ex\hbox{$\scriptscriptstyle#1$}}}}
    {\displaystyle{\mathop{\scriptscriptstyle\circ}_%
       {\scriptscriptstyle\raise.7ex\hbox{$\scriptscriptstyle#1$}}}}}}
\def\proj{\pi}
\def\anticomm#1,#2{\{#1,#2\}}
\def\maut{\alpha}
\def\mant{\beta}
\def\Pmau{P}
\def\id{\text{id}}
\def\Spsp{S}
\xdef\sp{{\lowercase{\Spsp}}}
\def\prmid{Q}
\def\gen{q}
\def\gene{p}
\def\Cg{\Gamma}
\def\Cac{\phi}
\def\CAC{\Phi}
\def\Csac{\psi}
\def\rep{\gamma}
\def\Rep{\Gamma}
\def\Carr{W}
\xdef\carr{\lowercase{\Carr}}
\def\acarr{z}
\def\End{\mathop{\rm End}}
\def\dmW{l}
\def\Ivs{R}
\def\lrep{\lambda}
\def\Id{J}
\def\Ib{b}
\def\rId{L}
\def\Intw{F}
\def\phm{\phantom{-}}
\def\sig#1{\ifcase#1 \or \sigma \or \epsilon \or \tau\fi}
\def\mat{M}
\def\indx{\nu}
\def\Slash#1{{#1\llap{/}}}
\def\Sslash#1#2{\Slash{{#1\mkern-#2mu}}\mkern#2mu}
\def\tr#1{{\text{tr}\left(#1\right)}}
\def\Re{\text{Re}\,}
\def\Im{\text{Im}\,}
\def\one{\mbox{\bf 1}}
\def\zero{\mbox{\bf 0}}
\def\Cl{{\cal C\it l}}
\def\adj#1{\overline{#1}}
\def\T#1{#1^T}

\def\mapsfrom{\leftarrow\mskip-4mu
  \vcenter{\hsize 0pt\llap{$\scriptscriptstyle\dashv$}}\mkern 4mu}
\def\mapsTo{\mkern 4mu\vcenter{\hsize 0pt\rlap{$\scriptscriptstyle\vdash$}}
  \mskip-4mu\rightarrow}
\let\Ch\check
\def\Cc{C}
\def\Spbl{B}
\def\Spal{E}
\def\Sphf{A}
\def\Calg{{\cal A}}
\def\Cbl{{\bf B}}
\def\Ctl{{\bf T}}
\def\fp{p}
\def\SC{\ifpreprintsty ;\\ \else ;\ \fi}
\def\acsz{\arraycolsep 0pt}
\begin{document}
\title{Octonionic representations of Clifford algebras and triality}
\author{J\"org Schray}
\address{Department of Physics, Oregon State University,
		Corvallis, OR 97331, USA \\
{\tt schrayj{\rm @}physics.orst.edu} }
\author{Corinne~A.~Manogue}
\address{Department of Physics, Oregon State University,
		Corvallis, OR  97331, USA \\
{\tt corinne{\rm @}physics.orst.edu}
}
\date{July 26, 1994}
\maketitle
\begin{abstract}
The theory of representations of Clifford algebras is extended to employ the
division algebra of the octonions or Cayley numbers.  In particular, questions
that arise from the non-associativity and non-commutativity of this division
algebra are answered.  Octonionic representations for Clifford algebras lead
to a notion of octonionic spinors and are used to give octonionic
representations of the respective orthogonal groups.  Finally, the triality
automorphisms are shown to exhibit a manifest $\perm_3 \times SO(8)$ structure
in this framework.
\end{abstract}


\section{Introduction}

The existence of classical supersymmetric string theories in $(n+1,1)$
dimensions has been linked to the existence of the normed division algebras
$\K_n$
\cite{Cl:supersymmetry and division algebras,Cl:Evans},  where $\K_n = \R$,
$\C$, $\H$, and $\O$ for $n = 1$, $2$, $4$, and $8$ are the algebras of the
reals, complexes, quaternions, and octonions.  One reason for this
correspondence is the isomorphism $sl(2,\K_n) \isom so(n+1,1)$ on the Lie
algebra level \cite{Cl:Sudbery}.
However, because of the non-associativity of the octonions,
the extension of this result to finite Lorentz transformations, i.e., on the
Lie group level, for $n=8$ has posed a problem until recently
\cite{Cl:Lorentz,Cl:Tachibana}.
Nevertheless, octonionic spinors based on
$sl(2,\O)$ have been used successfully as a tool to solve and parametrize
classical solutions of the superstring and superparticle
\cite{Cl:Tachibana,Cl:Manogue,Cl:Oda}.

Another link between octonions and supersymmetric theories is given by the
triality \cite{Cl:Adams,Cl:Evans,Cl:Porteous} automorphisms of $SO(8)$, which
interchange the spaces of vectors, even spinors, and odd spinors.  These
automorphisms are constructed using the Chevalley algebra, which combines
these three spaces into a single 24-dimensional algebra.  Our formulation of
the Chevalley algebra in terms of $3\times 3$ octonionic hermitian matrices
naturally extends to the exceptional Jordan algebra.  A variety of articles
connect this algebra to theories of the superstring, the superparticle, and
supergravity
\cite{Cl:Jordan,Cl:Jordanalg}.

Division algebras are also used in the spirit of GUTs to provide a group
structure that contains the known interactions \cite{Cl:Dixon}.

The contribution of this paper is to bring these many isolated observations
together and place them on the foundation of the theory of Clifford algebras.
Our framework allows an elegant unified derivation of all the previous results
about orthogonal groups.  The octonionic triality automorphisms, for example,
are completely symmetric with respect to the spaces of vectors, even spinors,
and odd spinors, as they should be.  We explain new features and
properties of octonionic representations of Clifford algebras related to the
possible choices of different octonionic multiplication rules.  We also find
that not all of the common constructions from complex representations have
exact analogues for octonionic representations because of the
non-commutativity of the octonions.  For example, the octonionic analogue of
the charge conjugation operation involves the opposite octonionic algebra,
without which the transformation behavior is inconsistent.  However, the extra
structure of two distinguished octonionic algebras may turn out to be a
feature of our formalism rather than a bug.

In a previous article
\cite{Cl:Lorentz} a demonstration of the construction of
$SO(7)$, $SO(8)$, $SO(9,1)$, and $G_2$ is given, which illustrates how the
octonionic algebra works explicitly.  However, in this article, we only use
the general algebraic properties of the octonions, rather than rely on
explicit computations involving a specific multiplication rule.  This approach
is taken to highlight the central role of the alternativity of the octonions
in the development of our formalism.  In essence, we suggest the division
algebra of the octonions not as an afterthought, but as a starting point for
incorporating Lorentzian symmetry and supersymmetry in supersymmetrical
theories.  This principle is brought to fruition in a fully octonionic
description of the triality automorphisms of the Chevalley algebra.

The content of this article is organized as follows: First we give a thorough
introduction to composition algebras and the division algebra of the
octonions.  In particular, we devote a large part of section
\ref{sec:Octonions} to the investigation of the relationship amongst different
multiplication tables of the octonions.
In section~\ref{sec:Clifford} we state basic concepts about Clifford algebras
and their representations.  We characterize the Clifford group and the
orthogonal group of a vector space with a metric by generating sets.  This
approach turns out to be better adapted to octonionic representations than
the usual Lie algebra one.
Then we introduce the octonionic representation of the Clifford algebra in
8-dimensional Euclidean space in section~\ref{sec:8}.  In section
\ref{sec:other reps}, the reductions to 7 and 6 dimensions and the extension to
9+1 dimensions are discussed.  In section~\ref{sec:triality}, we introduce an
octonionic description of the Chevalley algebra and show that the triality
symmetry is inherent in the octonionic description.  Then, in section
\ref{sec:full group}, we briefly explain how our results with regard to sets
of finite generators of Lie groups are related to the usual description in
terms of infinitesimal generators of the corresponding Lie algebra.  Section
\ref{sec:conclusion} discusses our results.

\section{The division algebra of the octonions}\label{sec:Octonions}

This section lays the first part of the foundation for octonionic
representations of Clifford algebras, namely it introduces the octonionic
algebra.
The first subsection deals with some general properties of composition
algebras.  A subsection introducing our convention for octonions follows.  We
then turn our attention to the relationship among different multiplication
tables for the octonions and introduce the opposite octonionic algebra.
For further information and omitted proofs see
\cite{Cl:Schafer,Cl:Zorn,Cl:Sudbery}.  A less rigorous approach is taken in
\cite{Cl:Lorentz}.

\subsection{Composition algebras}

An algebra $\alg$ over a field $\F$ is a vector space over $\F$ with a
multiplication that is distributive and $\F$-linear:
\begin{eqnarray}
\left.\acsz\begin{array}{rl}
x(y+z) ={}& xy + xz \\ \noalign{\smallskip} (x+y)z ={}& xz + yz
\end{array}
\right\}&\quad\forall\, x,y,z \in \alg,\\ \noalign{\medskip}
(fx)y = x(fy) = f(xy)& \qquad \forall\, x,y \in \alg, \>\forall\, f \in \F.
\label{eq:linear}
\end{eqnarray}
$\alg$ is also assumed to have a multiplicative identity $1_\alg$.

A composition algebra $\alg$ over a field $\F$ is defined to be an algebra
equipped with a non-degenerate symmetric $\F$-bilinear form,
\begin{equation}
\blmap : \vtop{\halign{$\hfil#\hfil$&$\hfil{}#{}\hfil$&$#\hfil$\cr
\alg \times \alg & \to & \F\cr\noalign{\smallskip}
(x,y) & \mapsto & \bilin x,y \,,\cr}}
\end{equation}
with the special property that it gives rise to a quadratic norm form which
is compatible with multiplication in the algebra:
\begin{eqnarray}
\Nmap : \vtop{\halign{$\hfil#\hfil$&$\hfil{}#{}\hfil$&$#\hfil$\cr
\alg & \to & \F\cr\noalign{\smallskip}
x & \mapsto & \normsq{x}:=\bilin x,x \,,\cr}}\\ \noalign{\bigskip}
\normsq{xy} = \normsq{x} \normsq{y} \quad \forall\, x,y \in \alg.
\label{eq:8squares}
\end{eqnarray}
(In the case of the octonions (\ref{eq:8squares}) is known as the
eight-squares theorem, i.e., a sum of eight squares is the product of two sums
of eight squares, and many applications rely on this identity.)
Two main consequences can be derived (see \cite{Cl:Schafer}) from this
essential
property of composition algebras.  Firstly, these algebras exhibit a weak form
of associativity:
\begin{equation}
\left.\acsz\begin{array}{rl}
x(xy) ={}& (xx)y \\ \noalign{\smallskip}
(yx)x ={}& y(xx)\end{array}
\right\}  \quad \forall\, x,y \in
\alg.\label{eq:alternative}
\end{equation}
Defining the associator as a measure of the deviation from associativity via
\begin{equation}
\assoc x,y,z := x(yz)-(xy)z, \qquad x,y,z \in \alg,
\end{equation}
then (\ref{eq:alternative}) implies
\begin{equation}
\assoc x,x,y = \assoc y,x,x = 0 \qquad \forall\, x,y \in \alg
\end{equation}
or (by polarization)
\begin{equation}
\assoc x,y,z = -\assoc x,z,y = -\assoc y,x,z  \qquad \forall\, x,y,z \in \alg,
\label{eq:alternating}\end{equation}
i.e., the associator is an alternating function of its arguments.  This weak
form of associativity is also called alternativity.  (\ref{eq:alternating})
and (\ref{eq:alternative}) are equivalent, if the characteristic $\ch{\F}$ of
$\F$ does not equal 2, which is assumed from now on.  As shown in
\cite{Cl:Schafer} alternativity implies the so-called Moufang \cite{Cl:Moufang}
identities,
\begin{equation}\left.\acsz\begin{array}{rl}
(xyx)z ={}& x(y(xz))\\ \noalign{\smallskip}
z(xyx) ={}& ((zx)y)x\\ \noalign{\smallskip}
x(yz)x ={}& (xy)(zx)
\end{array}\right\} \quad \forall\, x,y,z \in\alg,
\label{eq:Moufang}\end{equation}
which will turn out to be useful later on.

Secondly, composition algebras are endowed with an involutory antiautomorphism
$\Star$:
\begin{equation}\begin{array}{c}
\Star : \vtop{\halign{$\hfil#\hfil$&$\hfil{}#{}\hfil$&$#\hfil$\cr
\alg & \to & \alg\cr\noalign{\bigskip}
x & \mapsto & \Bar{x}:= 2\, \bilin 1,x - x\,,\cr}}\\ \noalign{\medskip}
\Bar{(xy)} = \Bar{y}\, \Bar{x} \qquad \forall\, x,y \in \alg.
\end{array}\label{eq:involution}\end{equation}
(Obviously, we view $\F$ as embedded in the algebra $\alg$ via $\F \isom
\F\,1_\alg \subseteq \alg$, in particular $1_\alg = 1_\F = 1$.  With this
identification and (\ref{eq:linear}), multiplication with an element of $\F$
is commutative, i.e., $\F \subseteq \Zc$, where $\Zc$ is the center of
$\alg$.)  We observe that $\Star$ is linear and fixes $\F$.  (Note that
$\bilin 1,1 = 1$, since $\bilin x,x = \bilin x,x \bilin 1,1 \quad \forall\, x
\in\alg.$)
This antiautomorphism can be shown to provide a way to express the quadratic
form $\Nmap$:
\begin{equation}
x\,\Bar{x} = \Bar{x}\,x = \normsq{x} \qquad \forall\, x \in \alg.
\label{eq:normsq}
\end{equation}
So all elements of $\alg$ satisfy a quadratic equation over $\F$:
\begin{equation}
x^2 -2\bilin 1,x x + \normsq{x} =0 \qquad\forall\, x \in \alg.
\label{eq:quadratic}
\end{equation}
Polarizing (\ref{eq:normsq}) results in an expression for the bilinear form:
\begin{equation}
\bilin x,y = {1 \over 2}(x\,\Bar{y} + y\,\Bar{x}) \qquad \forall\, x,y \in
\alg.
\end{equation}
We determine inverses:
\begin{equation}
x^{-1}={\Bar{x}\over \normsq{x}} \qquad \forall\, x \in \alg,\normsq{x} \neq 0.
\label{eq:inverse}
\end{equation}
However,
in order to solve a
linear equation $ax=b$, we need $a^{-1}(ax) = x$.  To see that we do indeed
have associativity in this case, we need the following relationship,
\begin{equation}
6 \assoc x,y,z
= \Comm x,{\comm y,z} + \Comm y,{\comm z,x} + \Comm z,{\comm x,y}
\qquad \forall\, x,y,z \in \alg,\label{eq:jacobi}
\end{equation}
between the associator and the commutator
\begin{equation}
\comm x,y := xy - yx, \qquad x,y \in \alg,
\end{equation}
which is defined as usual.
So for $\ch{\F} \neq 2,3$, we see that products with elements in $\Zc$ are
associative:
\begin{equation}
x \in \Zc \quad\; \iff \quad\; \comm x,y = 0 \quad \forall\, y \in \alg \quad\;
\implies \quad\; \assoc x,y,z = 0 \quad \forall\, y,z \in \alg.
\label{eq:F-assoc}
\end{equation}
Since the associator is linear in its arguments, we can put
(\ref{eq:inverse}), (\ref{eq:involution}), and (\ref{eq:F-assoc}) together:
\begin{equation}
\assoc x^{-1},x,y = {\assoc \Bar{x},x,y  \over \normsq{x}} = {2\bilin 1,x
\assoc 1,x,y - \assoc x,x,y \over \normsq{x}} = 0 \qquad \forall\, x,y \in
\alg,
\normsq{x} \neq 0.\label{eq:associative inverse}
\end{equation}
Finally, we observe more general consequences of (\ref{eq:involution}) and
(\ref{eq:F-assoc}):
\begin{eqnarray}
& \comm \Bar{x},y = -\comm x,y = \Bar{\comm x,y}
\qquad\forall\, x,y \in \alg\\
\noalign{\noindent and}
& \assoc\Bar{x},y,z = - \assoc x,y,z
= \Bar{\assoc x,y,z} \qquad\forall\, x,y,z \in \alg,
\end{eqnarray}
which imply that both commutators and associators have vanishing inner
products with 1:
\begin{equation}
\bilin 1,{\comm x,y} = \bilin 1,{\assoc x,y,z} = 0 \qquad\forall\, x,y,z \in
\alg.\label{eq:not real}
\end{equation}

We will now turn to the specific composition algebra of the octonions.

\subsection{Octonions}\label{subsec:octonions} 

According to a theorem by Hurwitz \cite{Cl:Hurwitz}, which relies heavily on
(\ref{eq:quadratic}) there are only four composition algebras over the reals
with a positive definite bilinear form, namely the reals, $\R$; the complexes,
$\C$; the quaternions, $\H$ \cite{Cl:Hamilton}; and the octonions or Cayley
numbers, $\O$ \cite{Cl:Cayley}. Their dimensions as vector spaces over $\R$
are 1, 2, 4, and 8.  Since the norm is positive definite, there exist inverses
for all elements except $0$ in these algebras.  Therefore, they are also
called normed division algebras.

For specific calculations the following concrete form of $\O$ is useful.  $\O
\isom \R^8$ as a normed vector space.  Fortunately, it is
always possible to choose an orthonormal basis $\{\iu_0$, $\iu_1$, $\ldots\,$,
$\iu_7\}$ which induces a particularly simple multiplication table for the
basis elements such as the one given by the following triples:
\begin{equation}\acsz\begin{array}{rll}
\iu_0 ={}& 1,\\
\iu_a^2 ={}& -1 \qquad (1\leq a\leq 7),\\ \noalign{\smallskip}
\iu_a\iu_b ={}& \iu_c = -\iu_b\iu_a \>\text{ and cyclic for }
\ifpreprintsty \\ \fi
(a,b,c) \in \Psym = \{(1,2,3), (1,4,5),& (1,6,7), (2,6,4), (2,5,7),
(3,4,7), (3,5,6)\}.
\end{array}\label{eq:basis definition}\end{equation}
The algorithm to obtain such a basis is similar to the Gram-Schmidt procedure
\cite{Cl:Linear Algebra} with additional requirements about products of
the basis elements (see \cite{Cl:Lorentz}).

Working over the field of real numbers, the following definitions of real and
imaginary parts are customary:
\begin{equation}\begin{array}{c}
\Re{x} := \bilin 1,x = {1\over 2}(x + \Bar{x}) \>\in \R,
\\ \noalign{\smallskip}
\Im{x} := x - \bilin 1,x = {1\over 2}(x - \Bar{x}) \>\in \R^\perp.
\end{array}\end{equation}
Also $\iu_0$ is called the real unit and the other basis elements are
called imaginary units,
\begin{equation}
\Re{\iu_0} = \iu_0, \qquad \Im{\iu_a} = \iu_a \qquad (1\leq a\leq 7).
\end{equation}
In analogy to $\C$ and $\H$, the antiautomorphism $\Star$ is called
``octonionic conjugation''.  it also changes the sign of the imaginary part.
With these conventions (\ref{eq:not real}) reads
\begin{equation}
\Re{\comm x,y} = \Re{\assoc x,y,z} = 0 \qquad\forall\, x,y,z \in \alg.
\label{eq:Re comm=0}
\end{equation}

\subsection{Multiplication tables}\label{subsec:mult tables}  

The question of possible multiplication tables arises, for example, when one
reads another article on octonions, which, of course, uses a different one
from the one given in (\ref{eq:basis definition}).  Usually it is remarked,
that all 480 possible ones are equivalent, i.e., given an octonionic algebra
with a multiplication table and any other valid multiplication table one can
choose a basis such that the multiplication follows the new table in this
basis.  One may also take the point of view, that there exist different
octonionic algebras, i.e., octonionic algebras with different multiplication
tables.  With this interpretation the previous statement means that all these
octonionic algebras are isomorphic.  However, this fact does not imply that a
physical theory might not make use of more than one multiplication table at
any given time.
In this section we extend the ideas of Coxeter \cite{Cl:Coxeter}, giving a
detailed description of how the various multiplication tables are related to
each other.
A new result, which emerges from our description, is that two classes of
multiplication tables can be identified, namely the class corresponding to a
given algebra and the one corresponding to its opposite algebra.
In a physical theory, the distinction between these two classes becomes
important when parity is not a good symmetry, i.e., in a chiral theory.

\begin{figure}
\ifpreprintsty\epsfysize=10.0cm\else\epsfysize=8.3cm\fi
\centerline{%
\epsffile[85 175 395 460]{Cfig1.eps}%
}
\caption{The projective plane $\plane$ representing a multiplication table for
the octonions.}
\label{triangle}
\end{figure}

The set
$\Psym$ in (\ref{eq:basis definition}) can be taken to represent a labeling of
the projective plane $\plane$ over the field with two elements $\Z2 = GF(2) =
\{ 0, 1 \}$ (see Fig.~\ref{triangle}).  Before we explain this correspondence,
we introduce the basic properties of $\plane$.  (Readers who are not familiar
with projective geometry may consult \cite{Cl:proj}.)  This plane contains as
points the one-dimensional linear subspaces of $(\Z2)^3$.
Given a basis of $(\Z2)^3$ these subspaces are
\begin{equation}\vcenter{\halign{$\hfil#\hfil$\cr
p_1 = \left\bra\left(\begin{array}{c} 1\\ 0\\ 0 \end{array}\right)\right\ket,
p_2 = \left\bra\left(\begin{array}{c} 0\\ 1\\ 0 \end{array}\right)\right\ket,
p_3 = \left\bra\left(\begin{array}{c} 1\\ 1\\ 0 \end{array}\right)\right\ket,
p_4 = \left\bra\left(\begin{array}{c} 0\\ 0\\ 1 \end{array}\right)\right\ket,
\cr
\noalign{\medskip}
p_5 = \left\bra\left(\begin{array}{c} 1\\ 0\\ 1 \end{array}\right)\right\ket,
p_6 = \left\bra\left(\begin{array}{c} 0\\ 1\\ 1 \end{array}\right)\right\ket,
p_7 = \left\bra\left(\begin{array}{c} 1\\ 1\\ 1 \end{array}\right)\right\ket.
\cr}}
\end{equation}
(Since these linear subspaces contain only one non-zero element, we will drop
the angle brackets and identify the points with the non-zero elements of
$(\Z2)^3$.)  The lines $l_1$, $l_2$, $\ldots\,$, $l_7$ of the plane are the
two-dimensional linear subspaces of $(\Z2)^3$, which can also be described by
their normal vectors $n_1$, $n_2$, $\ldots\,$, $n_7$, i.e., the dual vectors
that annihilate the subspaces:
\begin{equation}\ifpreprintsty\vcenter{\halign{$\hfil#\hfil$\cr \fi
n_1 = \left(\begin{array}{c} 1\\ 0\\ 0 \end{array}\right),
n_2 = \left(\begin{array}{c} 0\\ 1\\ 0 \end{array}\right),
n_3 = \left(\begin{array}{c} 1\\ 1\\ 0 \end{array}\right),
n_4 = \left(\begin{array}{c} 0\\ 0\\ 1 \end{array}\right),
\ifpreprintsty\cr
\noalign{\medskip}\fi
n_5 = \left(\begin{array}{c} 1\\ 0\\ 1 \end{array}\right),
n_6 = \left(\begin{array}{c} 0\\ 1\\ 1 \end{array}\right),
n_7 = \left(\begin{array}{c} 1\\ 1\\ 1 \end{array}\right).
\ifpreprintsty\cr}}\fi
\end{equation}
So there are also seven lines in $\plane$.  The geometry of the plane is then
defined by the incidence of points and lines, where
\begin{equation}
p_j \text{ and } l_k \text{ are incident}
\quad\iff\quad p_j \subset l_k
\quad\iff\quad \T{n_k} p_j \equiv 0 \pmod{2},
\end{equation}
for example, $p_3$, $p_5$, and $p_6$ are incident with $l_7.$

We are now in a position to specify the previously mentioned correspondence
between $\plane$ and $\Psym$.  $\Psym$ contains seven triples formed out of
seven labels.  The labels represent points and the triples represent lines
containing the three points given by the labels, i.e., a label and a triple
are incident, if and only if the label is part of the triple.  Cyclic
permutations of a triple change neither the multiplication table nor the
geometry of the plane.  However, $\Psym$ does
define an orientation on each line, since a transposition in a triple would
change the multiplication table.  This notion of orientation on the lines,
is represented by arrows in Fig.~\ref{triangle}.  So we can read the
multiplication table off the triangle.  If we follow a line connecting two
labels in direction of the arrow we obtain the product, for example, $i_3 i_4
= i_7$.
When moving opposite to the direction of the arrow we pick up a minus sign,
$i_4 i_2 = -i_6$.
(Note that in projective geometry the ends of the lines are connected, i.e.,
lines are topologically circles, $S^1$.

What are possible transformations of the multiplication table $\Psym$ and how
do they correspond to transformations of the projective plane $\plane$?
Looking at Fig.~\ref{triangle}, we see that there are three ways to change the
picture:
\begin{itemize}
\item[(i)] We may relabel the corners, leaving the arrows unchanged.
\item[(ii)] The labels may be kept fixed while some or all arrows are reversed.
\item[(iii)] Minus signs may be attached to the labels, i.e., we change part of
(\ref{eq:basis definition}) to read $\iu_a\iu_b= -\iu_c
= -\iu_b\iu_a$ and cyclic for $(a,b,c) \in \Psym$.
\end{itemize}
The sign change of a label in type~(iii) is equivalent to reversing the
orientation of the three lines through that point and therefore is included in
the transformations of type~(ii).  For the second kind of transformation, we
have to make sure, that the multiplication table so obtained satisfies
alternativity, for it to define another octonionic algebra.  One can show that
given the arbitrary orientation of four lines including all seven points, the
orientations of the remaining three lines are determined by alternativity.
(Note that there is only one case to consider.  Among the four lines there are
necessarily three which have one point in common.  Two of those together with
the fourth one fix one of the remaining orientations.)
%
%
%
This in turn implies that elementary transformations of type~(ii) change the
orientation of three lines which have one point in common.  So the
transformations of type~(ii) and type~(iii) are equivalent.  Since four arrows
can be chosen freely, we obtain sixteen as the number of possible
configurations of arrows, i.e., the number of distinct multiplication tables
that can be reached this way, namely: 1 original configuration with no
changes, 7 with the orientation of three lines through one point reversed, 7
with the orientation of four lines avoiding one point reversed, and 1 with the
orientation of all lines reversed.

In order to discuss these transformations further, we will introduce some
notation.  (Before developing this framework, I verified most of these results
using the computer algebra package Maple.  So the reader who is not
algebraically inclined may take this proof by exhaustion as sufficient.  For a
basic reference on group theory see \cite{Cl:group actions}.) We denote an
octonionic algebra given by an orthonormal basis of $\R^8$ and a set $\Psym$
of the type given in (\ref{eq:basis definition}) by $\O_\Psym$, and the set
made up of all such octonionic algebras by $\oset := \{\text{ all possible }
\Psym : \O_\Psym\}$.  ``All possible $\Psym$'' means those that induce a
multiplication table satisfying alternativity.  So $\oset$ can be viewed as
the set of possible multiplication tables.

We now consider the group action of $\tra = \tra_1 * \tra_2$, the free product
of transformations of type~(i) and (ii), on $\oset$:
\begin{equation}
\vtop{\halign{$\hfil#\hfil$&$\hfil{}#{}\hfil$&$#\hfil$\cr
\tra \times \oset & \to & \oset\cr
\noalign{\smallskip}
(\tsym,\O_\Psym) & \mapsto & \O_{\tsym(\Psym)} \,.\cr}}
\label{eq:action}\end{equation}
Thus each $\tsym \in \tra$ induces an isomorphism $\O_\Psym \mapright{\tsym}
\O_{\tsym(\Psym)}$.  The group of transformations $\tra_1$ of type~(i), i.e.,
the relabelings of the corners, is of course the permutation group on seven
letters, $\perm_7$, acting in the obvious way.  We identify the group $\tra_2$
of transformations of type~(ii) as $(\Z2)^7$, with the 7 generators acting as
the elementary transformations reversing the orientation of the three lines
through one point.  Earlier we saw that the orbits of an element of $\oset$
under the action of this group are of size 16:
$\order{\orbit{(\Z2)^7}{\O_\Psym}} = 16$.  In order to determine the orbits of
$\perm_7$ we first consider its subgroup $\sgr$ which acts as the group of
projective linear transformations on $\plane$ labeled as in
Fig.~\ref{triangle}, i.e., we let $\sgr$ act on one specific $\O_\Psym \in
\oset$, namely with $\Psym$ as in (\ref{eq:basis definition}).  $\sgr \isom
PGL(3,\Z2) \isom GL(3,\Z2)$ is generated by the permutations
$(1\,2\,4\,3\,6\,7\,5)$ and $(1\,2\,5)(3\,7\,4)$.  $\sgr$ is in fact simple, of
Lie-type, of order $168 = 2^3\cdot3\cdot7$, and denoted by $A_2(2)$ (see
\cite{Cl:simple groups}).  Since elements of $\sgr$ as projective linear
transformation do not change the geometry of $\plane$, they can only reverse
the orientations of lines, i.e., $\orbit{\sgr}{\O_\Psym} \subseteq
\orbit{(\Z2)^7}{\O_\Psym}$.
Hence, we have
$\order{\orbit{\sgr * (\Z2)^7}{\O_\Psym}} = 16$.
Thus the index of the
stabilizing subgroup of $\sgr$ has to divide 16:
\begin{equation}
[\sgr:\stab{\sgr}{\O_\Psym}] = \order{\orbit{\sgr}{\O_\Psym}}\; \Big|\; 16.
\end{equation}
Since the action of $\sgr$ is not trivial and $\sgr$ being simple of order 168
cannot have subgroups of index 2 or 4, we conclude
$\order{\orbit{\sgr}{\O_\Psym}} = 8$.  To determine
$\order{\orbit{\perm_7}{\O_\Psym}}$ we need to consider the cosets of $\sgr$ in
$\perm_7$.  There are  $[\perm_7:\sgr] = 30$ of them corresponding to distinct
geometries of $\plane$, i.e., the incidence of lines and points is different
for different cosets.  Therefore, there are 30 distinct classes of
multiplication tables, with members of one class related by a projective
linear transformation.  So it follows
\begin{equation}\acsz\begin{array}{rl}
\order{\orbit{\perm_7}{\O_\Psym}} ={}& 30\cdot 8 = 240,
\\ \noalign{\smallskip}
\order{\orbit{\tra}{\O_\Psym}} ={}& 30\cdot 16 = 480.
\end{array}\end{equation}
So relabelings of the corners reach only half of the possible multiplication
tables, which is a consequence of the fact that projective linear
transformations reach only half of the possible configurations of arrows.  Why
is this so and what are the possible implications?  To answer these questions
we need to understand how elements of $\sgr$ change orientations of lines.  We
can decompose the action of elements of $\sgr$ into one part that permutes the
lines and another one that reverses the orientation of certain lines in the
image.  An element $\tsym_1 \in \sgr$ of odd order $p$ may only change the
orientation of an even number of lines. For ${\tsym_1}^p = 1$ has to act
trivially on $\Psym$, and the changes of orientation add up modulo 2.
However, $\sgr$ is generated by elements of odd order, so all of its elements
change only the orientation of an even number of lines.  To obtain the full
orbit we may add just one element $\topp \in \tra_2$ that changes the
orientation of an odd number of lines.  A particularly good choice for $\topp$
is the product of all generators, i.e., the one corresponding to reversing all
seven lines (or attaching minus signs to all labels when viewed as type~(iii)
transformation).  Obviously, $\tsym_1\,\topp(P) = \topp\,\tsym_1(P) \quad
\forall \,\tsym_1 \in \tra_1$, so that we may form the direct product
${\tra_1}^\prime = \tra_1 \times \{1,\topp\}$ and
$\orbit{{\tra_1}^\prime}{\O_\Psym} = \orbit{\tra}{\O_\Psym}$.
Note that $\topp$ corresponds to the operation of octonionic conjugation, so
that the isomorphism given by $\topp$ is illustrated by the following diagram:
\begin{equation}
\begin{array}{ccc}
\O_\Psym \times \O_\Psym & \longrightarrow & \O_\Psym \\[5pt]
(a,b) & \longmapsto & ab\\[8pt]
\Bigr\downarrow\llap{$\vcenter{\hbox{\kern-3em $\topp \times \topp$}}$} &
\circlearrowleft &
\Bigl\downarrow\rlap{$\topp$}\\[8pt]
\O_{\topp(\Psym)} \times \O_{\topp(\Psym)} & \longrightarrow &
\O_{\topp(\Psym)}\\[5pt]
(\Bar{a},\Bar{b}) = (a^\prime,b^\prime) & \longmapsto
& \Bar{(ab)} = \Bar{b} \, \Bar{a} = b^\prime \, a^\prime
\end{array}\;.
\end{equation}
Therefore, $\O_{\topp(\Psym)}$ is the opposite algebra of $\O_\Psym$, i.e.,
the algebra obtained by reversing the order of all products.  So for
octonionic algebras, there is an isomorphism of an algebra and its opposite
algebra given by octonionic conjugation, besides the natural anti-isomorphism
given by identification.  What are the consequences of these results for a
physical theory?  Usually, the physical theory will contain a vector space of
dimension 8, for which we want to introduce an octonionic description. This
description, however, should be invariant under the appropriate symmetry
group, most commonly, $SO(8)$.  The multiplication table changes in a more
general way under $SO(8)$.  The product of two basis elements will turn out to
be a linear combination of all basis elements, but the relabelings given by
$\perm_7$ are certainly a subgroup contained in $SO(8)$.  Moreover, $\topp
\notin SO(8)$, which implies that the most general multiplication tables with
respect to an orthonormal basis split in two classes with $SO(8)$ acting
transitively on each class, but only $SO(8) \times \{1,\topp\} \isom O(8)$
acting transitively on all of them.  In fact we will find it useful to
consider two algebra structures, namely $\O$ and its opposite $\O_\opp$, on
the same $\R^8$ to describe the spinors of opposite chirality.

In a recent article, Cederwall \& Preitschopf \cite{Cl:S7} introduce an
``$X$-product'' on $\O$ via
\begin{equation}
a \cpr_X b := (a\,X)(\Bar{X}\,b), \qquad a,b,X \in \O,\; X\,\Bar{X} = 1,
\label{eq:x-product}
\end{equation}
which is just the original product for $X=1$.  As $X$ becomes different from
1, the multiplication table for this product changes continuously in a way
related to the $SO(8)$ transformations that leave $1$ fixed.  This changing
product appears naturally when the basis of a spinor space is changed, see
section~\ref{subsec:twisted spinors}.

\section{Clifford algebras and their representations}\label{sec:Clifford}

The second building block for octonionic representations of Clifford algebras
is presented in this section.
First we define an abstract Clifford algebra and observe some of its basic
properties.
Then we consider the Clifford group which
gives us the action of the orthogonal groups on vectors and spinors.  In our
approach to the Clifford group in this second subsection we also introduce the
key idea of characterizing groups by finite generators.
The
third subsection states the necessary facts about representations of Clifford
algebras, i.e., how we can find matrix algebras to describe Clifford algebras.
For further reference and proofs that are left out see \cite{Cl:Benn and
Tucker,Cl:Porteous,Cl:Lounesto,Cl:chessboard}. We only consider the real or
complex field, i.e., $\F = \R, \C$, in this section, even though some of the
statements generalize to other fields, in particular of characteristic
different from $2$.

\subsection{Clifford algebras}

The tensor algebra $\Tensor(V)$ of a vector space $V$ of dimension $\dmV$ over
a field $\F$ is the free associative algebra over $V$:
(All the products in this section are associative.)
\begin{equation}
\Tensor(V) := \oplus_{k=0}^{\infty} (V)^k,
\end{equation}
where
\begin{equation}
(V)^k = \underbrace{V \otimes V \otimes \cdots \otimes V}_{k \text{ copies}},
\;k>0,\quad (V)^0 = \F.
\end{equation}
The identity element is $1 \in \F$ and $\F$ lies in the center of
$\Tensor(V)$.
Given a metric $g$ on V, i.e., $g$ is a non-degenerate symmetric bilinear form
on $V$, the Clifford algebra $\Cl(V,g)$ is defined to be
\begin{equation}
\Cl(V,g) := {\quot \Tensor(V) / {I(g)}}\;,
\end{equation}
where
\begin{equation}
I(g) = {\bra u \otimes u - g(u,u):u \in V \ket}
\end{equation}
is the two-sided ideal generated by all expressions of the form $u \otimes u -
g(u,u)$.  If $V$ is unambiguously defined from the context, we simply write
$\Cl(g)$.  We denote multiplication in $\Cl(g)$ by
\begin{equation}
u \vm v := \proj^{-1}(u) \otimes \proj^{-1}(v) + I(g)
\qquad \forall \,u,v \in \Cl(g),
\end{equation}
where $\proj$ is the canonical projection:
\begin{equation}
\proj : \vtop{\halign{$\hfil#\hfil$&$\hfil{}#{}\hfil$&$#\hfil$\cr
\Tensor(V) & \to & \Cl(g)\cr
\noalign{\smallskip}
u & \mapsto & u +I(g)\cr}}
\end{equation}
and $\proj^{-1}(u)$ is any preimage of $u$.  Since $\proj$ restricted to $\F
\oplus V$ is injective, we identify this space with its embedding in
$\Cl(g)$.

{}From a more practical perspective a Clifford product is just a tensor product
with the additional rule that
\begin{equation}
u \vm u = g(u,u) \qquad \forall\, u \in V.\label{eq:vm}
\end{equation}
As a consequence elements of $V \subseteq \Cl(g)$ anticommute up to an
element of $\F$:
\begin{equation}
\anticomm u,v := u \vm v\, +\, v \vm u = 2 g(u,v) \qquad \forall\, u,v \in V
\label{eq:anticomm}
\end{equation}
or in terms of an orthonormal basis $\{\ob_1$, $\ob_2$, $\ldots\,$,
$\ob_\dmV\}$
\begin{equation}
\anticomm \ob_i,{\ob_j} := 2 g(\ob_i,\ob_j)
=
\left\lbrace\begin{array}{rl}
\pm 2, & \text{ for } i=j\\
0, & \text{ for } i \not= j
\end{array}\right.
\qquad (1 \leq i,j \leq \dmV).
\end{equation}
Based on these relationships, we find a basis for $\Cl(g)$ as a vector space,
\begin{equation}
\{ \ob_{a_1} \vm \ob_{a_2}\vm \cdots \vm \ob_{a_k} : 0\leq k\leq\dmV,\,
1 \leq a_1 < a_2 < \cdots \leq\dmV\},
\label{eq:Cl basis}
\end{equation}
which shows that
\begin{equation}
\dim \Cl(g) = \sum_{k=0}^\dmV {\dmV\choose k} = 2^\dmV.
\end{equation}
The product $\vol = \ob_1 \vm \ob_2 \vm \cdots \vm \ob_\dmV$ is called the
volume form and has the special property
\begin{equation}
\vol \vm u = (-1)^{\dmV+1} u \vm \vol \qquad \forall\, u \in V.\label{eq:vol}
\end{equation}
So for odd $\dmV$, $\vol$ lies in the center $\Zc$ of $\Cl(g)$.  In fact
\begin{equation}
\Zc =
\left\lbrace\begin{array}{ll}
\F, & \text{ for }\dmV\text{ even }\\
\F \oplus \F\vol,\; & \text{ for }\dmV\text{ odd }
\end{array}\right..
\end{equation}

There are two involutions on $\Tensor(V)$ given by the obvious extensions of
the following maps on $V\subseteq \Tensor(V)$: the main automorphism $\maut$,
\begin{equation}
\maut{\big|}_V : \vtop{\halign{$\hfil#\hfil$&$\hfil{}#{}\hfil$&$#\hfil$\cr
V & \to & V\cr
\noalign{\smallskip}
u & \mapsto & -u\,,\cr}}\label{eq:maut}
\end{equation}
and the main antiautomorphism $\mant$,
\begin{equation}
\mant{\big|}_V : \vtop{\halign{$\hfil#\hfil$&$\hfil{}#{}\hfil$&$#\hfil$\cr
V & \to & V\cr
\noalign{\smallskip}
u & \mapsto & u\cr
\noalign{\medskip}
& \hbox to 0pt{\hss $\mant(u \otimes v) = v \otimes u \quad \forall\,u,v\in
V\,.$\hss} & \cr
}}\label{eq:mant}
\end{equation}
Since $I(g)$ is invariant under $\maut$ and $\mant$, we obtain maps on the
quotient $\Cl(g)$.  The main antiautomorphism can also be understood as an
isomorphism between $\Cl(g)$ and its opposite algebra $(\Cl(g))_\opp$:
\begin{equation}
\begin{array}{ccc}
\Cl(g) \times \Cl(g) & \mapverylongright{\vm} & \Cl(g) \\[1ex]
(a,b) & \longmapsto & a \vm b\\[2ex]
\llap{$\mant \times \mant$}\Bigr\downarrow & \circlearrowleft &
\Bigl\downarrow\rlap{$\mant$}\\[2ex]
(\Cl(g))_\opp \times (\Cl(g))_\opp & \mapverylongright{\vm_\opp}
& (\Cl(g))_\opp\\[1ex]
(\mant(a),\mant(b)) = (a_\opp,b_\opp) & \longmapsto
& a_\opp \vm_\opp b_\opp = b_\opp \vm a_\opp\\[1ex]
&  & = \mant(b) \vm \mant(a) = \mant(a \vm b)
\end{array}\;.\label{eq:opposite algebra}
\end{equation}
The main automorphism $\maut$ defines a $\Z2$-grading on $\Cl(g)$ given by the
projections
\begin{equation}
\Pmau_0 := {1\over 2}(\id + \maut), \qquad \Pmau_1 := {1\over 2}(\id - \maut).
\end{equation}
The even and odd part of the Clifford algebra are defined to be
\begin{equation}
\Cl_0(g):= \Pmau_0(\Cl(g)), \qquad \Cl_1(g):= \Pmau_1(\Cl(g)).
\end{equation}
Then
\begin{equation}
\Cl(g) = \Cl_0(g) \oplus \Cl_1(g)
\end{equation}
and the even part $\Cl_0(g)$ is in fact a subalgebra of $\Cl(g)$.

We already saw how the Clifford algebra contains vectors.  A spinor space
$\Spsp$ is defined to be a minimal left ideal of $\Cl(g)$.  Such a space
\begin{equation}
\Spsp =  \Cl(g)\vm\prmid
\end{equation}
is generated by a primitive idempotent $\prmid\in\Cl(g)$, i.e.,
\begin{equation}
\prmid^2 = \prmid, \qquad
\Slash{\exists}\; \prmid_1,\prmid_2:\; \prmid_1^2 = \prmid_1 \not=0,\;
\prmid_2^2 = \prmid_2 \not=0,\;
\prmid = \prmid_1 + \prmid_2.
\end{equation}
(This characterization of minimal left ideals relies on the fact that Clifford
algebras over $\R$ and $\C$ are semisimple, see section~\ref{subsec:reps}.)
If the primitive idempotent is even, the spinor space $\Spsp$ decomposes into
the spaces of even and odd Weyl spinors:
\begin{equation}
\Spsp = \Spsp_0 \oplus \Spsp_1, \qquad \Spsp_k = \Pmau_{k} \Spsp =
\Cl_k(g)\vm\prmid,\; (k=0,1).\label{eq:Weyl spinors}
\end{equation}
Different names for these spaces are used within the mathematical physics
community.  $\Spsp$ is also called the space of Dirac spinors and $\Spsp_0$
and $\Spsp_1$ are called semi-spinor spaces.  Sometimes, elements of $\Spsp$
are called bi-spinors and elements of $\Spsp_0$ and $\Spsp_1$ are just called
even and odd spinors.

For a mixed primitive idempotent $\prmid$ there may
still be a Weyl decomposition (see (\ref{eq:Weyl projection})), but it is not
compatible with the $\Z2$ grading on $\Cl(g)$:
\begin{equation}
\Spsp = \Cl(g)\vm\prmid = \Cl_0(g)\vm\prmid = \Cl_1(g)\vm\prmid.
\end{equation}

For odd $\dmV$, $\Spsp$ is also called the space of Pauli spinors or
semi-spinors.  If only the double $2\Spsp := \Spsp \oplus \Spsp$ carries a
faithful representation of $\Cl(g)$ (see section~\ref{subsec:reps}), then some
authors refer only to $2\Spsp$ as the space of spinors.

\subsection{The Clifford group}\label{subsec:Cg}

The connection of the symmetry group of the metric, i.e., the orthogonal
group, with the Clifford algebra is made in this subsection via the Clifford
group $\Cg(g)$ (see (\ref{eq:hom Cg O even}), (\ref{eq:hom Cg O odd}), and
(\ref{eq:hom Cg0 SO})).  We use a non-standard definition of the Clifford
group in terms of a set of finite generators.  We were led to this approach
because octonionic representations are naturally implemented in this way.
However, we feel this characterization of the Clifford group is simpler in
many applications.  By relating both definitions to the orthogonal group we
show that they are essentially equivalent (see (\ref{eq:Cg prime even}),
(\ref{eq:Cg prime odd}), and (\ref{eq:Cg0 prime})).

We define the Clifford group $\Cg(g)$ to be the group generated by
the vectors of non-zero norm, i.e.,
\begin{equation}
\Cg(g) := \bra u \in V \subseteq \Cl(g) : u^2 = g(u,u) \not= 0 \ket.
\label{eq:def1 Cg}
\end{equation}
As we will see, this definition is almost equivalent to the usual one,
\begin{equation}
\Cg^\prime(g) := \{ u \in \Cl(g) : u {\rm\ invertible},\, u \vm x \vm u^{-1}
\in V \;\; \forall\, x \in V \} \supseteq \Cg(g).
\label{eq:def2 Cg}
\end{equation}
Considering $u \in\Cg(g)\cap V$ and any $x \in V$ we see that
\begin{equation}\vcenter{\halign{$\hfil#$&$#\hfil$\cr
u \vm x \vm u^{-1} ={}& u \vm x \vm {1\over g(u,u)} u
= {1\over g(u,u)} (- x \vm u + 2g(x,u)) \vm u \cr \noalign{\smallskip}
={}& - x + 2{g(x,u)\over g(u,u)} u \in V.\cr}}
\label{flip}
\end{equation}
Therefore, $\Cg^\prime(g)\supseteq \Cg(g)$ indeed, and in particular
$\Cg^\prime(g)\cap V = \Cg(g)\cap V$.  In fact, the definition of
$\Cg^\prime(g)$ implies that $\Cg(g)$ is stable under conjugation in
$\Cg^\prime(g)$, i.e., $\Cg(g)$ is a normal subgroup of $\Cg^\prime(g)$.  We
will investigate the structure of the Clifford group on the basis of this
group action of $\Cg^\prime(g)$ on $V$:
\begin{equation}
\Cac^\prime:\vtop{\halign{$\hfil#$&$\hfil{}#{}\hfil$&$#\hfil$\cr
\Cg^\prime(g)\times V & \to & V\cr
\noalign{\smallskip}
(u,x) & \mapsto & \Cac^\prime_u(x) := u\vm x\vm u^{-1}\,.\cr}}
\label{eq:Cp action}
\end{equation}
Dropping all the primes we have the obvious restriction
\begin{equation}
\Cac:\vtop{\halign{$\hfil#$&$\hfil{}#{}\hfil$&$#\hfil$\cr
\Cg(g)\times V & \to & V\cr
\noalign{\smallskip}
(u,x) & \mapsto & \Cac_u(x) := u\vm x\vm u^{-1}\,.\cr}}\label{eq:tensor action}
\end{equation}
(We will not explicitly give the unprimed analogues of expressions below.)
Of course, these actions can be extended to give inner automorphisms of
$\Cl(g)$.
According to (\ref{flip}), the action of $u \in V \cap \Cg^\prime(g)$
is just a reflection of $x$ at the hyperplane orthogonal to $u$ composed
with an inversion of the whole space.
In particular $\Cac^\prime_u(x) \in V$ and $\Cac^\prime_u$ is an isometry:
\begin{equation}\acsz\begin{array}{rl}
g(\Cac^\prime_u(x),\Cac^\prime_u(x))
={}&\Cac^\prime_u(x) \vm \Cac^\prime_u(x)
= ({1\over u^2})^2 u \vm x \vm u^{-1} \vm u \vm x \vm u^{-1}\\[1ex]
={}& u \vm x \vm  x \vm u^{-1} = g(x,x).
\end{array}\end{equation}
So (\ref{eq:Cp action}) (resp.\ (\ref{eq:tensor action})) gives a homomorphism
$\CAC^\prime$ (resp.\ $\CAC$) of $\Cg^\prime(g)$ (resp.\ $\Cg(g)$) to the group
of isometries or orthogonal transformations $O(g)$ of $V$:
\begin{equation}
\CAC^\prime:\vtop{\halign{$\hfil#$&$\hfil{}#{}\hfil$&$#\hfil$\cr
\Cg^\prime(g) & \to & O(g)\cr
\noalign{\smallskip}
u & \mapsto &
\Cac^\prime_u:\vtop{\halign{$\hfil#$&$\hfil{}#{}\hfil$&$#\hfil$\cr
V & \to & V\cr
\noalign{\smallskip}
x & \mapsto & \Cac^\prime_u(x)=u\vm x\vm u^{-1}\cr}}\cr}}
\end{equation}
To compare $\Cg^\prime(g)$ (resp.\ $\Cg(g)$) with $O(g)$ we need to know the
range and the kernel  of $\CAC^\prime$ (resp.\ $\CAC$).
Since the reflections at hyperplanes generate all orthogonal transformations
$\CAC^\prime$ (resp.\ $\CAC$) is onto, if we can find a preimage of the
inversion $x\mapsto -x$.
Because of (\ref{eq:vol}), $\vol \in \Cg(g) \subseteq \Cg^\prime(g)$ does the
job for even $\dmV$.
For odd $\dmV$, there is no element of $\Cl(g)$ that anticommutes with all
$x\in V$. So there is no preimage of the inversion, which leaves us with
$SO(g)$ as the range.  The kernel coincides with the part of the center, that
lies in the Clifford group.
Thus we have according to the homomorphism theorems
\begin{eqnarray}
& \quot\Cg(g)/{\F^{*}} \isom  O(g) \isom \quot\Cg^\prime(g)/{\F^{*}}
& \qquad\text{(for even\ }\dmV)\label{eq:hom Cg O even}\\
\noalign{\medskip}
& \quot\Cg(g)/{\F^{*}\bra\vol\ket} \isom SO(g)  \isom
\quot\Cg^\prime(g)/{\Zc^{*}}
& \qquad\text{(for odd\ }\dmV),\label{eq:hom Cg O odd}
\end{eqnarray}
where $\F^{*} = \F \setminus \{0\}$, $\bra\vol\ket$ is the group generated by
$\vol$, and $\Zc^{*}:=\Cg^\prime(g)\cap \Zc$ is the invertible part of the
center.  So the Clifford group is isomorphic to the orthogonal (resp.\ simple
orthogonal) group up to a subgroup of the center $\Zc$.
Therefore,
\begin{eqnarray}
& \Cg(g) \isom \Cg^\prime(g)
& \qquad \text{(for even\ }\dmV)\label{eq:Cg prime even}\\
\noalign{\medskip}
& \Cg(g) \times\quot\Zc^{*}\!/{\F^{*}\bra\vol\ket} \isom \Cg^\prime(g)
& \qquad \text{(for odd\ }\dmV),\label{eq:Cg prime odd}
\end{eqnarray}
So for even $\dmV$ both definitions (\ref{eq:def1 Cg}) and (\ref{eq:def2 Cg})
of the Clifford group are equivalent.  For odd $\dmV$ they differ by
inhomogeneous elements of the invertible part of the center $\Zc^{*}$.  For
our purposes it will be sufficient to consider the Clifford group $\Cg(g)$ as
defined in (\ref{eq:def1 Cg}) only.

For both even and odd $\dmV$, we obtain a homomorphism from the even Clifford
group $\Cg_0(g)$,
\begin{equation}
\Cg_0(g) := \Cg(g) \cap \Cl_0(g) = \Pmau_0\Cg(g) = \Cg^\prime(g) \cap \Cl_0(g)
= \Pmau_0\Cg^\prime(g),\label{eq:Cg0 prime}
\end{equation}
onto $SO(g)$:
\begin{equation}
\quot\Cg_0(g)/{\F^{*}} \isom SO(g).\label{eq:hom Cg0 SO}
\end{equation}
The even Clifford group is generated by pairs of vectors with non-zero norm:
\begin{equation}
\Cg_0(g) = \bra u \vm v : u,v \in V,\; g(u,u) \not= 0 \not= g(v,v) \ket.
\end{equation}
In fact one of the vectors may be fixed,
\begin{equation}
\Cg_0(g) = \bra u \vm w : u \in V,\; g(u,u) \not= 0 \ket,\;
\text{ for some }w \in V,\,g(w,w) \not= 0,
\label{eq:one vector fixed}
\end{equation}
since any product of two vectors $u,v$ can be written as a product of two
pairs that contain $w$:
$u \vm v = (u \vm w) \vm (v^\prime \vm w)$, where $v^\prime = {1\over g(w,w)}
w \vm v \vm w^{-1} \in V$.

We also have an action $\Csac$ of $\Cg(g)$ on the Clifford algebra $\Cl(g)$
and in particular on any of its minimal left ideals, a space of spinors
$\Spsp$:
\begin{equation}
\Csac:\vtop{\halign{$\hfil#$&$\hfil{}#{}\hfil$&$#\hfil$\cr
\Cg(g)\times \Spsp & \to & \Spsp\cr
\noalign{\smallskip}
(u,\sp) & \mapsto & \Csac_u(\sp) :=  u \vm \sp\,.\cr}}
\label{eq:spinor action}
\end{equation}

So we have two actions of the Clifford group $\Cg(g)$ and its subgroup
$\Cg_0(g)$, the action $\Cac$ on vectors (\ref{eq:tensor action}) and the
action $\Csac$ on spinors (\ref{eq:spinor action}).  These actions give rise
to the so-called vector and spinor representations of the simple orthogonal
group via the isomorphism (\ref{eq:hom Cg0 SO}).  All octonionic
representations of orthogonal groups in sections~\ref{sec:8} and
\ref{sec:other reps} are based on this relationship.
In physics particles are understood in terms of representations of groups
describing the symmetries in the physical theory, in particular the Lorentz
group.  Therefore these representations of the orthogonal group are important,
because they determine how physical fields transform.

The way in which $\F^*$ should be divided out in (\ref{eq:hom Cg0 SO}) is
obvious for the vector representation, since $\F^*$ is the kernel of $\CAC$.
For the spinor representation, requiring the invariance of the spinor bilinear
form (see section \ref{sec:bilinear forms}) determines how to divide out
scalars (see (\ref{eq:Cg normalized}) and (\ref{eq:Cg normalized})).
Actually, this leads to a homomorphism of $\Cg_0(g)$ onto the universal
covering group of $SO(g)$, which is also called $Spin(g)$.  We will take
$SO(g)$ to be the appropriate group depending on the context and not make a
distinction in notation between $SO(g)$ and $Spin(g)$.

\subsection{Representations of Clifford algebras}\label{subsec:reps}

In this subsection we describe how we can get a matrix algebra that is
isomorphic to a Clifford algebra.  In a sense this is the analogue to
\ref{subsec:octonions}, where we gave an explicit form of the octonions, which
implemented their abstract properties.  We start out by introducing some
definitions concerning representations in general.  Algebras are assumed to be
finite dimensional and contain a unit element. (For a general reference for
representation theory see, e.g., \cite{Cl:Curtis and Reiner}.)

A representation $\rep$ of an algebra $\alg$ over a field $\F$ in a
vector space $\Carr$ is a homomorphism
\begin{equation}
\rep:\vtop{\halign{$\hfil#$&$\hfil{}#{}\hfil$&$#\hfil$\cr
\alg & \to & \End_\F(\Carr)\cr
\noalign{\smallskip}
a & \mapsto & \rep(a):\vtop{\halign{$\hfil#$&$\hfil{}#{}\hfil$&$#\hfil$\cr
\Carr & \to & \Carr\cr
\noalign{\smallskip}
\carr & \mapsto & \rep(a)\carr\,,\cr}}\cr}}
\end{equation}
i.e.,
\begin{equation}
\left.\acsz\begin{array}{rl}
\rep(a \vm b) ={}& \rep(a) \rep(b)\\ \noalign{\smallskip}
\rep(a + b) ={}& \rep(a) + \rep(b)
\end{array}\right\}
\qquad \forall\, a,b \in \alg,
\end{equation}
where we denote multiplication in $\alg$ by $\vm$ even though $\alg$ is
not necessarily a Clifford algebra.
Given a basis of $\Carr$, $\rep(a)$ as an endomorphism of $\Carr$ may be
understood as an $\dmW \times \dmW$-matrix, where $\dmW = \dim W$ is called
the dimension of the representation.  The representation is called faithful,
if $\rep$ is injective.  $\Ivs$ is an invariant subspace of $\rep$, if
$\rep(a)\Ivs \subseteq \Ivs \quad \forall a\in\alg$.
The representation $\rep$ is called irreducible, if there are no invariant
subspaces of $\rep$ other than $\Carr \not= \{0\}$ and $\{0\}$.
A reducible representation $\rep$ may be reduced to a representation
$\rep_\Ivs$ on an invariant subspace $\Ivs$, i.e., $\alg \mapright{\rep_\Ivs}
\End_\F(\Ivs)$ requiring $\rep_\Ivs(a)\carr = \rep(a)\carr \quad \forall\,
\carr \in \Ivs,\; a \in \alg$.
An algebra is called simple, if it allows a faithful and irreducible
representation.  An algebra is called semisimple if it is a direct sum of
simple algebras.

Since a left ideal $\Id$ of $\alg$ is by definition stable under left
multiplication,
\begin{equation}
\alg \vm \Id \subseteq \Id,
\end{equation}
and since $\Id$ is a vector space, we have a natural representation
$\lrep_\Id$ of $\alg$ on $\Id$.
(Again given a basis $\{\Ib_1$, $\Ib_2$, $\ldots\,$, $\Ib_\dmW\}$ we have a
representation in terms of matrices: $a \vm \Ib_i = \lrep_\Id(a)_{ij}
\Ib_j$.)
Taking $\Id = \alg$ we obtain the so called left regular representation,
which is faithful.  If $\Id$ is a minimal left ideal, then the representation
on it is irreducible, since invariant subspaces would correspond to proper
subspaces of $\Id$ which are left ideals and contradict the minimality of
$\Id$.

If the algebra $\alg$ is semisimple then the converse is also true, i.e., any
irreducible representation can be written as a $\lrep_\Id$ for some minimal
left ideal $\Id$:  In this case an irreducible representation $\rep$ of $\alg =
\alg_1 \oplus \alg_2 \oplus \cdots \oplus \alg_k$ is an
irreducible representation of one of the simple components, say $\alg_k$.  So
a minimal ideal $\rId$ of $\rep(\alg)$ can be lifted to a minimal ideal
of $\Id \subseteq \alg_j$, such that $\rep(\Id) = \rId$.  Then the following
diagram commutes,
\begin{equation}
\begin{array}{ccc}
\alg & \mapverylongright{\rep} & \rep(\alg) \subseteq \End(W)\\[1ex]
\llap{$\lrep_\Id$}\Bigr\downarrow & \circlearrowleft
& \Bigl\downarrow\rlap{$\lrep_\rId$}\\[2ex]
\End(\Id) & \mapverylongright{\rep(\Id)=\rId} & \End(\rId)\\
\end{array}\;.
\end{equation}
Since the maps $\lrep_\rId$ and $\rep(\Id)=\rId$ are isomorphisms, there is an
isomorphism relating $\Carr$ and $\Id$ as vector spaces,
\begin{equation}
\Intw:\Carr \to \Id,
\end{equation}
such that
\begin{equation}
\rep(a) \circ \Intw = \Intw \circ \lrep_\Id(a) \qquad \forall\, a \in \alg.
\label{eq:equi reps}
\end{equation}
$\Intw$ is said to intertwine the representations $\rep$ and $\lrep_\Id$:
\begin{equation}
\begin{array}{ccc}
\Carr & \mapverylongright{\rep(a)} & \Carr \\[1ex]
\llap{$F$}\Bigr\downarrow & \circlearrowleft & \Bigl\downarrow\rlap{$F$}\\[2ex]
\Id & \mapverylongright{\lrep_\Id} & \Id\\
\end{array}\quad.
\end{equation}
Representations related in this way are called equivalent.  In terms of their
matrix form, equivalent representations are related by a basis transformation.
This observation also shows that for a simple algebra all irreducible
representations are equivalent to $\lrep_\Id$ and therefore equivalent to each
other.

As is shown in the references given (see in particular \cite{Cl:Benn and
Tucker,Cl:chessboard}), Clifford algebras over $\R$ and $\C$ are simple or
semisimple. Therefore, there is an equivalent definition for spinors in terms
of representations of $\Cl(g)$, i.e., a spinor space $\Spsp$ can be defined to
be the carrier space of an irreducible representation of $\Cl(g)$.

In order to find a concrete representation, we must still find a primitive
idempotent $\prmid$ that generates a minimal left ideal $\Id$ and observe how
the basis elements of $\Cl(g)$ act on it.  Actually, we will give a procedure
to construct a representation that does not use a primitive idempotent
explicitly.  For this purpose we define the signature of a metric for the case
$\F=\R$.  We say that $g$ has the signature $p,q$ (written $g_{p,q}$), where
$\dim V = p + q = \dmV$, if there is an orthonormal basis $\{\ob_1$, $\ob_2$,
$\ldots\,$, $\ob_\dmV\}$ of $V$, such that
\begin{equation}
g_{ij} := g(\ob_i,\ob_j) =
\left\lbrace\begin{array}{rl}
0, & \text{\ for\ }i \not= j\\
1, & \text{\ for\ }i = j \leq p\\
-1, & \text{\ for\ }i = j > p
\end{array}\right..\label{eq:signature}
\end{equation}
We write $\Cl(p,q)$ and $\rep_{p,q}$ to
denote $\Cl(g_{p,q})$ and one of its representations.  It is particularly
simple to give a procedure that produces a representation of
$\Cl(\dm2V,\dm2V)$, i.e., in the case of a so-called neutral space.  The
procedure starts by ``guessing'' a representation $\rep_{1,1}$ for $\Cl(1,1)$:
\begin{equation}\acsz\begin{array}{rl}
&\rep_{1,1}(\ob_1) 
:= \pmatrix{0 & 1 \cr 1 & 0\cr} =: \sig1
\quad\text{\ and\ }\quad
\rep_{1,1}(\ob_2)
:= \pmatrix{ 0 & -1 \cr 1 & \phm0 \cr} =: \sig2
\\ \noalign{\medskip}
\implies\quad{}&
\rep_{1,1}(\ob_1 \vm \ob_2)
= \pmatrix{ 1 & \phm0 \cr 0 & -1 \cr} =: \sig3
\quad\text{\ and\ }\quad
\rep_{1,1}(1)
= \pmatrix{ 1 & 0 \cr 0 & 1 \cr} = \one.
\end{array}\label{eq:rep cl11}\end{equation}
Notice that the representation is completely specified by defining it on a
basis of $V$, since $V$ generates the algebra.  In order to ensure that these
assignments actually lead to a representation of the Clifford algebra, we need
to check that (\ref{eq:anticomm}) is satisfied for all pairs of images of
basis elements, i.e.,
\begin{equation}
\anticomm \rep(\ob_i),{\rep(\ob_j)}
= \rep(\ob_i)\,\rep(\ob_j) + \rep(\ob_j)\,\rep(\ob_i)
= 2 g_{ij}\,\one \qquad (1 \leq i,j \leq \dmV).
\label{eq:rep check}
\end{equation}
The representation $\rep_{1,1}$ is faithful and irreducible, since its image
is the space $\mat_2(\F)$ of $2\times2$-matrices.  So there are no proper
invariant subspaces and the dimensions of $\Cl(1,1)$ and $\mat_2(\F)$ match.
This representation may be used as a building block to give the Cartan
extension of a faithful
irreducible representation $\rep_{p,q}$ of $\Cl(p,q)$, ($p+q = 2\dm2V$ even)
to a representation $\rep^\prime$ of $\Cl(V^\prime,g^\prime)$ with $\dim
V^\prime = 2\dm2V +2$:
\begin{equation}\ifpreprintsty\vcenter{\halign{$\hfil#\hfil$\cr \fi
\rep^\prime(\ob_i^\prime) = \sig1 \otimes  \rep_{p,q}(\ob_i)
\qquad(1\leq i\leq2\dm2V),
\ifpreprintsty\cr
\noalign{\medskip}
\else
\qquad
\fi
\rep^\prime(\ob_{2\dm2V+1}^\prime) = \sig1 \otimes  \rep_{p,q}(\vol),
\qquad
\rep^\prime(\ob_{2\dm2V+2}^\prime) = \sig2 \otimes  \rep_{p,q}(1).
\ifpreprintsty\cr}}\fi
\label{eq:rep extension}\end{equation}
(Of course, there are other extensions using the same building blocks.)  It
is easy to check that $\rep^\prime$ is faithful and irreducible if
$\rep_{p,q}$ was.  The signature of the resulting metric $g^\prime$ depends on
the value of
\begin{equation}
\rep_{p,q}(\vol)^2 = (-1)^{{1\over 2}\indx(\indx -1)}\,\one,
\end{equation}
where $\indx := p-q = 2(p-m) = 2(m-q)$ is called the index of the metric
$g_{p,q}$.  So for even (resp.\ odd) ${\indx\over 2}$, we obtain a
representation $\rep^\prime$ of $\Cl(p+1,q+1)$ (resp.\ $\Cl(p,q+2)$).  Since
for neutral spaces $\indx = 0$, we can get any $\rep_{\dm2V,\dm2V}$, starting
from $\rep_{1,1}$ by iteration of this extension.  We note that the dimension
$\dmW$ of the carrier space of this irreducible representation is
$2^\dm2V=2^{\dmV \over 2}$.

For even ${\indx\over 2}$, $\rep_{p,q}(\vol)$ has eigenvalues $+1$ and $-1$
and we have Weyl projections $\Pmau_\pm$ (\ref{eq:Weyl spinors}):
\begin{equation}
\Pmau_\pm := {1\over 2}(1 \pm \vol).
\label{eq:Weyl projection}
\end{equation}
One of these projectors can be decomposed to give an even primitive idempotent
$\prmid$.
A representation such as the one given, where
$\rep_{p,q}(\vol) = \left({
{\bf 1}_{\dm2V\times\dm2V}\;\phm0_{\phantom{\dm2V\times\dm2V}} \atop
0_{\phantom{\dm2V\times\dm2V}} \; -{\bf 1}_{\dm2V\times\dm2V}}\right)$
is called a Weyl representation, since the Weyl projections $\Pmau_\pm$ take a
simple form.  Due to the property (\ref{eq:vol}) of $\vol$,
\begin{equation}
\Pmau_\pm a = a_0 \Pmau_\pm + a_1 \Pmau_\mp,
\end{equation}
where $a_0$ and $a_1$ are the even and odd part of $a$.  Since either
$\Pmau_+$ or $\Pmau_-$ annihilates the even primitive idempotent $\prmid$, we
indeed get projections onto the spaces of even and odd Weyl spinors.  Let, for
example, $\Pmau_+ \vm \prmid = \prmid$ and $\Pmau_- \vm \prmid = 0$, then for
$\sp = a \vm \prmid = a_0 \vm \prmid + a_1 \vm
\prmid \in \Spsp_0 \oplus \Spsp_1$, $a = a_0 + a_1$ as before,
\begin{equation}\acsz\begin{array}{rl}
\Pmau_+ \vm \sp ={}& \Pmau_+ \vm a \vm \prmid
= a_0 \vm \Pmau_+ \vm \prmid + a_1 \vm \Pmau_- \vm \prmid = a_0 \vm \prmid \in
\Spsp_0,\\ \noalign{\smallskip}
\Pmau_- \vm \sp ={}& \Pmau_- \vm a \vm \prmid
= a_0 \vm \Pmau_- \vm \prmid + a_1 \vm \Pmau_+ \vm \prmid
= a_1 \vm \prmid \in \Spsp_1.
\end{array}\end{equation}
If we choose a mixed primitive idempotent $\prmid$, then we get a different
decomposition $\Spsp = \Pmau_+ \Spsp \oplus \Pmau_- \Spsp$ unrelated to the
decomposition of the Clifford algebra in its even and odd part.

The procedure continues for even $\dmV$ and $\indx\not=0$.  In this
case we can get a complex representation of the same dimension $\dmW =
2^\dm2V$ by complexifying and transforming the metric to obtain a neutral
space.
(We can change the sign of $g(\ob_j,\ob_j)$ for given $j$ using
the transformation $\ob_k \mapsto
\left\lbrace
{\phantom{i}\ob_k, \text{\ for\ } k\not=j
\atop
i\ob_j, \text{\ for\ } k=j}\right.$.
Thus we may choose a basis to obtain a form (\ref{eq:signature}) of the metric
with any $p,q$ where $p+q=\dmV$.)
This complex representation is faithful and irreducible but not
necessarily equivalent to a real one.  To examine this issue we define the
complex conjugate $\Bar{\rep}$ of a representation $\rep :
\alg \to \End_\C(\Carr)$ by
\begin{equation}
\Bar{\rep}: \vtop{\halign{$\hfil#$&$\hfil{}#{}\hfil$&$#\hfil$\cr
\alg & \to & \End_\C(\Bar{\Carr})\cr
\noalign{\smallskip}
a & \mapsto & \Bar{\rep}(a)=\Bar{(\rep(a))}\,.\cr}}
\end{equation}
If $\alg$ is simple then $\rep$ and $\Bar{\rep}$ are equivalent, i.e., there
exists a linear map $\Cc:\Carr\to\Bar{\Carr}$ intertwining these two
representations:
\begin{equation}
\Bar{\rep}(a) \circ \Cc= \Cc \circ \rep(a) \qquad \forall\,a\in\alg.
\label{eq:charge conj}
\end{equation}
It follows by complex conjugation that
\begin{equation}
\rep(a)\circ\Bar{\Cc} \circ \Cc
= \Bar{\Cc} \circ \Bar{\rep}(a) \circ \Cc
= \Bar{\Cc}\circ \Cc \circ \rep(a)\qquad \forall\,a\in\alg,
\label{eq:Schur}
\end{equation}
whence by Schur's Lemma $\Bar{\Cc}\circ \Cc$ is proportional to the identity.
Since $\Bar{\Cc}\circ \Cc$ has a real eigenvalue, $\Cc$ can be
normalized to satisfy
\begin{equation}
\Bar{\Cc}\circ \Cc = \pm \one.
\end{equation}
If and only if $\Bar{\Cc}\circ \Cc = +\one$, then we can find a basis
transformation to make $\rep_{p,q}$ real.  This is the case for $\indx \equiv
0,2 \pmod{8}$.  In practice, we relate $\Carr$ and $\Bar{\Carr}$ by complex
conjugation in the obvious way.  $\Cc$ is found by imposing (\ref{eq:charge
conj}) for $a \in \{\ob_1$, $\ob_2$, $\ldots\,$, $\ob_\dmV\}$.  (Following the
procedure given above, any of the matrices $\rep(\ob_k)$ is either real or
purely imaginary, so that $\Cc$ either commutes or anticommutes with it.)  The
new basis is a basis of eigenvectors for $\Cc$, which is invariant under $\sp
\mapsto \sp_\Cc:=\Bar{(\Cc\,\sp)}$.  ($\sp_\Cc$ is essentially the charge
conjugate spinor for $\sp$.)
For the cases $\indx \equiv 0,6 \pmod{8}$ we can make a
similar transformation to make $\rep_{p,q}$ purely imaginary.  These real
(resp.\ purely imaginary) representations are known as Majorana representations
of the first (resp.\ second) kind.  Of course, even for $\indx \equiv 4,6
\pmod{8}$ we can find an irreducible real representation of higher dimension,
namely $\dmW = 2^{\dm2V +1}$, by letting
$1 \to \left({1\;0\atop0\;1}\right) = \one$
and
$i \to \left({0\;-1\atop1\;\phm0}\right) = \sig2$
in an irreducible complex representation.

{}From a faithful, irreducible representation $\rep$ of the full Clifford
algebra $\Cl(g)$ we derive a representation $\rep_0$ of the even subalgebra
$\Cl_0(g)$ by the obvious restriction.  $\rep_0$ is faithful, but not
irreducible, except for real representations when $\indx\equiv 2\pmod{8}$.
For $\indx\equiv 0,4 \pmod{8}$, there are two-sided ideals of $\Cl_0(g)$
generated by the idempotents ${1\over 2}(1\pm\vol)$.
Each of these two-sided ideals $\Id$ carries an irreducible representation of
dimension $2^{\dm2V-1}$, but only the double $2\Id$ carries a faithful
representation.  For $\indx\equiv 6 \pmod{8}$ the isomorphism (\ref{eq:even
subalgebra}) in the following paragraph shows that $\Cl_0(q,p) \isom
\Cl_0(p,q)$, hence we know the dimension of the irreducible representation to
be $\dmW = 2^\dm2V$ from the case $\indx\equiv 2\pmod{8}$.

Representations $\rep_{p,q}$ with odd $\dmV$ can be obtained by shrinking a
representation of higher dimension, since we have the isomorphisms
\begin{eqnarray}
& \Cl_0(q+1,p) \isom \Cl(p,q) \isom \Cl_0(p,q+1) &
\label{eq:even subalgebra}\\
\noalign{\noindent obtained from extending}
& \ob_1\ob_{k+1} \mapsfrom \ob_k \mapsTo \ob_k
\ob_{\dmV+1} & \qquad(1\leq k\leq\dmV).
\label{eq:shrink rep}
\end{eqnarray}
Given the procedure above we can find an irreducible representation of
$\Cl(p,q)$ by constructing one corresponding to an even subalgebra for even
$\dmV$.
According to the isomorphism (\ref{eq:even subalgebra}) which also holds true
for $p+q$ even, we can shrink representations for odd $\dmV$ in a similar way.

Irreducible representations of the Clifford algebra $\Cl(g)$ induce
irreducible representations of the Clifford group $\Cg(g)$, since the basis
elements of $\Cl(g)$ as in (\ref{eq:Cl basis}) are contained in $\Cg(g)$.  The
representations arising from the tensor (resp.\ spinor) action (\ref{eq:tensor
action}) (resp.\ (\ref{eq:spinor action})) are known as the vector (resp.\
spinor) representation of $\Cg(g)$.

\subsection{Bilinear forms on spinors}\label{sec:bilinear forms}

Physical observables are tensors, which in terms of the Clifford algebra
transform under the orthogonal group like (\ref{eq:tensor action}), while
spinors transform like (\ref{eq:spinor action}).  For this reason it seems
likely that a bilinear form on spinors may provide observables based on
spinors.  The algebraic approach uses the fact that for $u\in\Cg(g)$ its
inverse $u^{-1}$ is proportional to $\mant(u)$.  Therefore, up to a
normalization $\sp\mant(\sp^\prime)$ transforms under the tensorial action of
$\Cg(g)$.  A decomposition in terms of a basis of the Clifford algebra gives
the tensorial pieces of certain rank.  In terms of representations we
construct a bilinear form on spinors considering induced representations of
the opposite Clifford algebra.  Given a representation
$\rep:\alg\to\End_\F(\Carr)$ there is an induced representation $\T\rep$, its
``transpose'':
\begin{equation}
\T\rep: \vtop{\halign{$\hfil#$&$\hfil{}#{}\hfil$&$#\hfil$\cr
\alg_\opp & \to & \End_\F(\T\Carr)\cr
\noalign{\smallskip}
a_\opp & \mapsto & (\T{\rep})(a_\opp):=\T{(\rep(a))}
:\vtop{\halign{$\hfil#$&$\hfil{}#{}\hfil$&$#\hfil$\cr
\T{\Carr} & \to & \T{\Carr}\cr
\noalign{\smallskip}
\T\carr & \mapsto & \T{\rep}(a_\opp)(\T{\carr})=\T{\carr}\rep(a_\opp).
\cr}}
\cr}}
\end{equation}
This is indeed a representation since
\begin{equation}\acsz\begin{array}{rl}
\T\rep(a_\opp \vm_\opp b_\opp)
={}& \T{(\rep(b_\opp \vm a_\opp))} = \T{(\rep(b)\rep(a))}
\\ \noalign{\smallskip}
={}& \T\rep(a_\opp)\T\rep(b_\opp) \qquad \forall\,a_\opp,b_\opp\in\alg_\opp.
\end{array}\end{equation}
As we pointed out in (\ref{eq:opposite algebra}), the main antiautomorphism
$\mant$ can be viewed as connecting the algebra $\alg$ and its opposite
$\alg_\opp$, so that we may obtain another induced representation $\Ch\rep$
for $\alg$ by
\begin{equation}
\Ch\rep(a) := \T\rep(\mant(a)) = \T{(\rep(\mant(a)))}
\qquad (a\in\alg),
\end{equation}
where we interpret $\mant$ first $\alg \maplongright{\mant} \alg_\opp$ as in
(\ref{eq:opposite algebra}) and then as an antiautomorphism $\alg
\maplongright{\mant} \alg$ on $\alg$.

Since a bilinear form on spinors can be understood as a linear transformation
$\Spbl:\Carr \to \T\Carr$, we take $\Spbl$ to be a map that intertwines
the representations $\rep$ and $\Ch\rep$.
Such a map exists if the representation $\rep$ is irreducible, whence
$\Ch\rep$ is also irreducible.  In this case $\Spbl$ is defined up to a
constant by
\begin{equation}
\Spbl\circ\rep(a)=\Ch\rep(a) \circ \Spbl \quad \forall\,a \in \alg \qquad
\Longleftrightarrow\qquad
\Spbl \rep(\ob_k)= \T{(\rep(\ob_k))} \Spbl
\quad \forall\,k\in\{1,\ldots,\dmV\}.
\end{equation}
We understand $\Spbl$ as a bilinear form on $\Carr$:
\begin{equation}
\Spbl: \vtop{\halign{$\hfil#$&$\hfil{}#{}\hfil$&$#\hfil$\cr
\Carr\times\Carr & \to & \F\cr
\noalign{\smallskip}
(\sp,\sp^\prime) & \mapsto &
\Spbl(\sp,\sp^\prime):=(\Spbl(\sp))(\sp^\prime) = \adj\sp \sp^\prime
= \T{\sp}\Spbl\sp^\prime
\cr}}
\end{equation}
both as a map and as its matrix form.  $\adj\sp:= \Spbl(\sp) = \T{\sp}\Spbl$
is the adjoint to $\sp$ with respect to $\Spbl$.
Indeed, $\Spbl(\sp,\sp^\prime)$ transforms
like a scalar (compare (\ref{eq:spinor action})):
\begin{equation}\acsz\begin{array}{rl}
\Spbl(\sp,\sp^\prime) \mapsrightto{\psi_u}
\Spbl(u\vm\sp,u\vm\sp^\prime)
={}& \T{\sp} \T{\rep(u)} \Spbl \rep(u) \sp^\prime
=\T{\sp} \Ch\rep(\mant(u)) \Spbl \rep(u) \sp^\prime
\\ \noalign{\smallskip}
={}& \T{\sp} \Spbl \rep(\mant(u)) \rep(u) \sp^\prime
= [\mant(u)\vm u]\> \T{\sp} \Spbl \sp^\prime,
\end{array}\end{equation}
if $u = u_1 \vm \cdots \vm u_k \in \Cg(g)$ and $u_1$, $\ldots\,$, $u_k \in V$
such that
\begin{equation}
\mant(u)\vm u = g(u_1,u_1)\ldots g(u_k,u_k) = 1.\label{eq:Cg normalized}
\end{equation}
For $x \in V$, $x \vm
\sp^\prime \mapsrightto{\psi_u} u \vm x \vm \sp^\prime = (u \vm x \vm u^{-1})
\vm (u \vm \sp^\prime)$, hence $\Spbl(\sp,x\vm\sp^\prime)$ also transforms
like a scalar.  Therefore, a vector $y$ is given by
\begin{equation}
y_k = \Spbl(\sp,\ob_k\vm\sp^\prime) = \T\sp \Spbl \rep(\ob_k) \sp^\prime
\qquad (1\leq k\leq\dmV).
\end{equation}
In a similar way, a tensor $Y$ of rank r may be formed:
\begin{equation}
Y_{k_1\ldots k_r}
= \Spbl(\sp,\ob_{k_1}\vm\cdots\vm\ob_{k_r}\vm\sp^\prime)
= \T\sp \Spbl \rep(\ob_{k_1})\ldots\rep(\ob_{k_r}) \sp^\prime
\qquad(1\leq k_1,\ldots,k_r\leq\dmV).
\label{eq:tensor bilinear}
\end{equation}

Another bilinear form $\Spal$ may be obtained by replacing the main
antiautomorphism $\mant$ with $\maut\circ\mant$ which, of course, is an
antiautomorphism also.  So $\Spal$ is determined up to a constant by
\begin{equation}
\Spal\circ\rep(a)=\Ch\rep(\maut(a)) \circ \Spal
\quad \forall\,a \in \alg
\ifpreprintsty
\quad
\else\qquad\fi
\Longleftrightarrow
\ifpreprintsty
\quad\!
\else\qquad\fi
\Spal \rep(\ob_k)=  - \T{(\rep(\ob_k))} \Spal
\quad \forall\,k\in\{1,\ldots,\dmV\};
\end{equation}
therefore, for even $\dmV$
\begin{equation}
\Spal = \Spbl \rep(\vol).
\end{equation}
The condition (\ref{eq:Cg normalized}) changes to
\begin{equation}
(\maut\circ\mant)(u)\vm u = (-1)^k g(u_1,u_1)\ldots g(u_k,u_k) = 1,
\label{eq:Cg Enormalized}
\end{equation}
which reduces to the previous condition for $u\in\Cg_0(g)$.
So both bilinear forms are invariant under the action of normalized elements
of $\Cg_0(g)$.

Both of these bilinear forms may be combined with $\Cc$ to give a sesquilinear
form $\Sphf:\Carr \to \Carr^\dagger$ on $\Carr$.
We only consider the combination $\Sphf := \Bar{\Spbl} \circ \Cc$ here:
\begin{equation}\acsz\begin{array}{rrl}
&\Sphf \circ \rep(a)
={}& \Bar{\Spbl} \circ \Cc \circ \rep(a)
= \Bar{\Spbl} \circ \Bar{\rep}(a) \circ \Cc
= \T{\Bar{\rep}(\mant(a))} \circ \Bar{\Spbl} \circ \Cc
\\ \noalign{\smallskip}
&={}& \rep^\dagger(\mant(a)) \circ \Sphf \qquad \forall\,a \in \alg.
\\ \noalign{\medskip}
\Longleftrightarrow\quad&\Sphf \rep(\ob_k) ={}&\rep^\dagger(\ob_k) \Sphf
\qquad \forall\,k\in\{1,\ldots,\dmV\},
\label{eq:sesquilinear form}
\end{array}\end{equation}
By a similar argument as in (\ref{eq:Schur}),
\begin{equation}\acsz\begin{array}{rl}
(\Sphf^{-1})^\dagger \circ \Sphf \circ \rep(a)
={}& (\Sphf^{-1})^\dagger \circ \rep^\dagger(\mant(a)) \circ
(\Sphf^\dagger \circ (\Sphf^{-1})^\dagger) \circ \Sphf
\\ \noalign{\smallskip}
={}& (\Sphf^{-1})^\dagger \circ (\Sphf \circ \rep(\mant(a))^\dagger
\circ (\Sphf^{-1})^\dagger \circ \Sphf
\\ \noalign{\smallskip}
={}& (\Sphf^{-1})^\dagger \circ
(\rep^\dagger((\mant\circ\mant)(a)) \circ \Sphf)^\dagger
\circ (\Sphf^{-1})^\dagger \circ \Sphf
\\ \noalign{\smallskip}
={}& ((\Sphf^{-1})^\dagger \circ \Sphf^\dagger) \circ
\rep(a) \circ \Sphf^\dagger \circ \Sphf
\\ \noalign{\smallskip}
={}& \rep(a) \circ \Sphf^\dagger \circ \Sphf \qquad \forall\,a \in \alg,
\end{array}\end{equation}
we conclude by Schur's Lemma that we can normalize $\Sphf$ to satisfy
\begin{equation}
(\Sphf^{-1})^\dagger \circ \Sphf = \one.
\end{equation}
Therefore, $\Sphf$ may be assumed to be hermitian.  Of course, $\Sphf$ like
$\Spbl$ may be used to define a spinor adjoint $\adj\sp := \Sphf(\sp) =
\sp^\dagger\Sphf$ and to construct tensors of various rank as sesquilinear
forms of spinors.  Which one of these forms is chosen depends on the signature
and the physical theory.

In all of our derivations involving $\Cc$, $\Spbl$, $\Spal$, and $\Sphf$, we
relied on certain properties of matrix multiplication over the field $\C$
(resp.\ $\R$), namely the fact that transposition is an antiautomorphism and
complex conjugation is an automorphism of matrix multiplication.  We are about
to replace $\F$ by $\O$.
Since octonionic multiplication is not commutative and octonionic conjugation
has become an antiautomorphism, the only remaining antiautomorphism of
octonionic matrix multiplication is hermitian conjugation.  Due to the
non-associativity of the octonions even the carrier space $\Carr$ is no longer
a vector space, but an ``octonionic module''.  It is surprising but true that
there are natural resolutions for these difficulties as we show in the
following section \ref{sec:8}.

\section{An octonionic representation of ${\cal C%
\it\noexpand\noexpand\noexpand\lowercase{l}}(8,0)$}\label{sec:8}

In this section we will put the results of sections~\ref{sec:Octonions} and
\ref{sec:Clifford} to work and examine the features of octonionic
representations of Clifford algebras, considering the example of $\Cl(8,0)$.
So $V=\R^8$ with a positive definite norm.  Let $\{\ob_0$, $\ob_1$,
$\ldots\,$, $\ob_7\}$ be an orthonormal basis of $V$.
Note that we choose indices ranging from 0 to 7 in this section.  The
octonionic algebra $\O$ is assumed to be given with basis $\{\iu_0$, $\iu_1$,
$\ldots\,$, $\iu_7\}$ obeying the multiplication table (\ref{eq:basis
definition}).  However, the properties
\begin{equation}\acsz\begin{array}{rll}
\iu_0 ={}& 1, \\ \noalign{\smallskip}
\iu_a^2 ={}& -1 & \qquad (1\leq a\leq 7),\\ \noalign{\smallskip}
\iu_a\iu_b ={}& -\iu_b\iu_a & \qquad (1\leq a < b \leq 7)
\end{array}\end{equation}
rather than the particular multiplication rule, i.e., the particular set
$\Psym$ of triples, will be relevant.
Furthermore, we identify $V$ with $\O$ as vector spaces by $x^k\ob_k\mapsto
x^k\iu_k$.

\subsection{The representation}

An octonionic representation $\rep_{8,0}: \Cl(8,0) \to \mat_2(\O)$ is given by
\begin{eqnarray}
\rep_{8,0}(\ob_k)&:=&%
\pmatrix{ 0 & \iu_k \cr\Bar{\iu_k} & 0 \cr} =:\Rep_k \qquad (0\leq k \leq 7)
\\ \noalign{\medskip}
\iff \qquad%
\rep_{8,0}(x)&:=&\pmatrix{ 0 & x \cr\Bar{x} & 0 \cr}
= x^k\Rep_k =: \Slash{x}\qquad (x \in V).
\label{eq:8orep}
\end{eqnarray}
The carrier space $\Carr$ of the representation is understood to be $\O^2$,
i.e., the set of columns of two octonions, with $\rep_{8,0}(x)$ acting on it
by left multiplication.  Therefore, octonionic matrix products are interpreted
as being associated to the right and acting on $\Carr$, i.e., octonionic
matrix multiplication is understood to be composition of left multiplication
onto $\Carr$.  For example, if we want to verify that (\ref{eq:8orep}) is a
representation, then checking that
\begin{equation}
\Slash{x}\Slash{x}
= \pmatrix{ 0 & x \cr \Bar{x} & 0 \cr}
\pmatrix{ 0 & x \cr \Bar{x} & 0 \cr}
= \pmatrix{ x\,\Bar{x} & 0 \cr 0 & \Bar{x}\,x \cr}
= \normsq{x} \one = g(x,x)\one \qquad\forall\,x\in V
\end{equation}
in accordance with (\ref{eq:rep check}) is not sufficient.  This relationship
has to hold even when acting on an element $\carr = \left( \carr_0 \atop
\carr_1 \right) \in \Carr$:
\begin{equation}\acsz\begin{array}{rl}
\Slash{x}\Slash{x}\carr
:={}& \pmatrix{0 & x\cr\Bar{x} & 0\cr}
\bigl(\pmatrix{0 & x\cr\Bar{x} & 0\cr}\pmatrix{\carr_0\cr\carr_1\cr}\bigr)
= \pmatrix{x\,(\Bar{x}\carr_0)\cr \Bar{x}(x \carr_1) \cr}
= \pmatrix{(x\,\Bar{x})\carr_0\cr (\Bar{x}x) \carr_1 \cr}
\\ \noalign{\smallskip}
={}& \normsq{x} \carr \qquad
\forall\,\carr\in\Carr\;\forall\,x\in V.
\end{array}\end{equation}
Thus the alternative property (\ref{eq:associative inverse}) of the octonions
ensures the validity of the representation.

For the representation to be irreducible, we need to show that there are no
non-trivial invariant subspaces.  We do this in two steps.  First, we show
that $\left({1 \atop 0}\right) \in \Carr$ can be mapped to any $\carr \in
\Carr$:
\begin{equation}
(\Sslash{\Bar{\carr_1}}{8} + \Sslash{\carr_0}{8} \Slash{1})\pmatrix{1\cr0\cr}
= \pmatrix{0 & \Bar{\carr_1}\cr \carr_1 & 0\cr}\pmatrix{1\cr0\cr}
+ \pmatrix{0 & \carr_0\cr\Bar{\carr_0} & 0\cr}
\bigl(\pmatrix{0 & 1\cr1 & 0\cr}\pmatrix{1\cr0\cr}\bigr)
= \pmatrix{\carr_0\cr \carr_1\cr}.
\end{equation}
Second, we will show that any non-zero $\carr\in\Carr$ can be mapped to
$\left({1\atop 0}\right)$, using the Weyl projections $\Pmau_\pm$.
If this is so, then there are no non-trivial invariant subspaces of the
representation $\rep_{8,0}$.

Since (\ref{eq:vol}) holds for the volume element $\vol$, we have for $\Rep_9
:= \rep_{8,0}(\vol) =
\Sslash{\iu_0}{4}\Sslash{\iu_1}{4}\ldots\Sslash{\iu_7}{4}$
\begin{equation}\acsz\begin{array}{rl}
&\phm\Rep_9 \Slash{x}
= \pmatrix{
0 & \iu_0(\Bar{\iu_1}(\iu_2(\ldots(\Bar{\iu_7} x)\ldots)))\cr
\Bar{\iu_0}(\iu_1(\Bar{\iu_2}(\ldots(\iu_7 \Bar{x})\ldots))) & 0\cr}
\\ \noalign{\smallskip}
={}& -\Slash{x} \Rep_9
= \pmatrix{
0 & -x(\Bar{\iu_0}(\iu_1(\Bar{\iu_2}(\ldots(\Bar{\iu_6} \iu_7)\ldots))))\cr
-\Bar{x}(\iu_0(\Bar{\iu_1}(\iu_2(\ldots(\iu_6\Bar{\iu_7} )\ldots)))) & 0\cr}
\,\quad\forall\,x \in V,
\end{array}\end{equation}
hence
\begin{equation}
\iu_0(\iu_1(\iu_2(\ldots(\iu_7 x)\ldots)))
= x(\iu_0(\iu_1(\iu_2(\ldots(\iu_6 \iu_7)\ldots)))) \qquad \forall\,x\in\O.
\end{equation}
Since $\Rep_9^2=1$, $\Rep_9$ has eigenvalues $\pm 1$, whence we can find
solutions to the equation
\begin{equation}\begin{array}{c}
\Rep_9 \carr = \pm \carr
\\ \noalign{\medskip\smallskip}
\Longleftrightarrow\quad\!
\pmatrix{
\Bar{\iu_0}(\iu_1(\Bar{\iu_2}(\ldots(\Bar{\iu_7} \carr_0)\ldots)))\cr
\iu_0(\Bar{\iu_1}(\iu_2(\ldots(\iu_7 \carr_1)\ldots)))\cr}
= \pmatrix{
-\carr_0(\Bar{\iu_0}(\iu_1(\Bar{\iu_2}(\ldots(\Bar{\iu_6} \iu_7)\ldots))))\cr
-\carr_1(\iu_0(\Bar{\iu_1}(\iu_2(\ldots(\iu_6\Bar{\iu_7} )\ldots))))\cr}
= \pm \pmatrix{\carr_0\cr\carr_1\cr}.
\end{array}\end{equation}
Since a non-trivial solution exists,
\begin{equation}
\iu_0(\iu_1(\iu_2(\ldots(\iu_7 x)\ldots))) = \pm x \qquad \forall\,x\in\O.
\label{eq:complete product}
\end{equation}
Which sign is true depends on the specific multiplication rule.  With our
convention the plus sign applies.  In fact, the sign difference corresponds to
the two classes of multiplication tables.  Since $\Rep_9$ is defined by its
action under left multiplication, we have an octonionic Weyl representation:
\begin{equation}
\Rep_9 = \pmatrix{1 & 0\cr0 & -1\cr}.
\end{equation}
The Weyl projections take the form
\begin{equation}
\Pmau_+ = \pmatrix{1 & 0\cr0 & 0\cr},\qquad
\Pmau_- = \pmatrix{0 & 0\cr0 & 1\cr}.
\end{equation}
For any $0\not=\carr\in\Carr$, at least one of $\Pmau_+\carr$ or
$\Pmau_-\carr$ does not vanish.  If $\Pmau_+\carr\not=0$, then
\begin{equation}
\Sslash{\carr_0}{8}^{-1}\Slash{1}\Pmau_+\carr
=\pmatrix{0 & \carr_0^{-1}\cr\Bar{(\carr_0^{-1})} & 0\cr}
\pmatrix{0 & 1\cr1 & 0\cr}
\pmatrix{1 & 0\cr0 & 0\cr}\pmatrix{\carr_0\cr\carr_1\cr}
= \pmatrix{1\cr0\cr}.
\end{equation}
(Note that $\Slash{1} = \Rep_0$ corresponds to a vector $\ob_0 \in V \subseteq
\Cl(8,0)$ and is to be distinguished from the identity $\rep(1) = \one$.)
If $\Pmau_-\carr\not=0$, then
\begin{equation}
\Sslash{\carr_1}{8}^{-1}\Pmau_-\carr
=\pmatrix{0 & \carr_0^{-1}\cr\Bar{(\carr_0^{-1})} & 0\cr}
\pmatrix{0 & 0\cr0 & 1\cr}\pmatrix{\carr_0\cr\carr_1\cr} = \pmatrix{1\cr0\cr}.
\end{equation}
This completes the proof that $\rep_{8,0}$ is irreducible.  Since $\Cl(8,0)$
is simple, it does not contain any two-sided ideals other than $\{0\}$ and
itself, which are also the only candidates for the kernel of any
representation of $\Cl(8,0)$.  Therefore, $\rep_{8,0}$ is faithful, since it
is not trivial.  Faithfulness of the representation can also been shown
constructively without using the fact that $\Cl(8,0)$ is simple.  One has to
check, for example, if the dimension of the algebra generated by $\{\Rep_0$,
$\Rep_1$, $\ldots\,$, $\Rep_7\}$ is $2^8$.  Another approach is to construct
orthogonal transformations (see \cite{Cl:Lorentz}), since the Clifford group
spans the Clifford algebra.  So if the representation obtained for the
Clifford group is faithful, then so is the representation for the Clifford
algebra.

In this article we have chosen to rely only on the algebraic properties of the
octonions, rather than using the correspondence to a real representation.
However, for completeness, we give the matrices corresponding to left
multiplication with respect to our convention:
\begin{equation}\acsz\begin{array}{rlrl}
\Rep_0 ={}& \sig1 \otimes \one \otimes \one \otimes \one, &\qquad
\Rep_1 ={}& -\sig2 \otimes \one \otimes \one \otimes \sig2,
\\ \noalign{\smallskip}
\Rep_2 ={}& -\sig2 \otimes \sig3 \otimes \sig2 \otimes \sig3, &\qquad
\Rep_3 ={}& -\sig2 \otimes \one \otimes \sig2 \otimes \sig1,
\\ \noalign{\smallskip}
\Rep_4 ={}& -\sig2 \otimes \sig2 \otimes \one \otimes \sig3, &\qquad
\Rep_5 ={}& -\sig2 \otimes \sig2 \otimes \sig3 \otimes \sig1,
\\ \noalign{\smallskip}
\Rep_6 ={}& -\sig2 \otimes \sig1 \otimes \sig2 \otimes \sig3, &\qquad
\Rep_7 ={}& -\sig2 \otimes \sig2 \otimes \sig1 \otimes \sig1.
\end{array}\label{eq:rep8 explicit}\end{equation}

Since we have an irreducible representation, we may identify the carrier
space $\Carr$ with the space of spinors.  So for now we consider elements of
$\O^2$ as octonionic spinors.  Later in section~\ref{subsec:twisted spinors}
we will add a subtle twist to this understanding.

\subsection{The hermitian conjugate representation and spinor covariants}
\label{subsec:herm conj rep}

Since octonionic conjugation is an antiautomorphism of $\O$, the octonionic
conjugate of the product of two matrices is not the product of the octonionic
conjugates.  Matrix transposition requires a commutative multiplication to be
an antiautomorphism.  Thus only hermitian conjugation, which combines both
operations, remains as an antiautomorphism of $\mat_2(\O)$.  More precisely,
for products of three matrices we need to keep the grouping of the product the
same, i.e., under hermitian conjugation left multiplication by a matrix goes
to right multiplication by its hermitian conjugate and vice versa.  So we can
define $\Ch\rep_{8,0}: \Cl(8,0) \to (\mat_2(\O))^\dagger$ by
\begin{equation}
\Ch\rep_{8,0}(a) := (\rep_{8,0}(\mant(a)))^\dagger \qquad (a \in \Cl(8,0)).
\end{equation}
This representation acts on the set $\Carr^\dagger =(\O^2)^\dagger$ of rows of
two octonions by right multiplication.  It is also faithful and irreducible
and therefore equivalent to $\rep_{8,0}$.  The isomorphism $\Sphf$
intertwining $\rep_{8,0}$ and $\Ch\rep_{8,0}$ is given by
\begin{equation}
\Sphf:\vtop{\halign{$\hfil#\hfil$&$\hfil{}#{}\hfil$&$\hfil#\hfil$\cr
\Carr & \to & \Carr^\dagger\cr
\noalign{\smallskip}
\carr=\left( \carr_0 \atop \carr_1 \right) & \mapsto &
\carr^\dagger = (\Bar{\carr_0},\Bar{\carr_1}).
\cr}}
\end{equation}
Its matrix form is just the identity,
\begin{equation}
\Sphf = \pmatrix{1 & 0\cr 0 & 1\cr},
\end{equation}
which is verified,
\begin{equation}
\Sphf \circ \rep_{8,0}(a) = \Ch\rep_{8,0}(a) \circ \Sphf
\quad \forall\,a\in\Cl(8,0)
\ifpreprintsty
\quad\!
\else\qquad\fi
\Longleftrightarrow
\ifpreprintsty
\quad\!
\else\qquad\fi
\Sphf \rep_{8,0}(x) = (\rep_{8,0}(x))^\dagger \circ \Sphf
\quad \forall\,x\in V,
\end{equation}
considering $\Rep_k = (\Rep_k)^\dagger \quad (0\leq k \leq 7)$.

{}From $\Sphf$ we obtain a hermitian form on $\Carr$:
\begin{equation}
\Sphf:\vtop{\halign{$\hfil#$&$\hfil{}#{}\hfil$&$#\hfil$&$#\hfil$\cr
\Carr\times\Carr & \to & \R\cr
(\carr,\acarr) & \mapsto &
\Sphf(\carr,\acarr)
:={}&(\Sphf(\carr))(\acarr) = \Re \carr^\dagger\Sphf\acarr
= \Re (\Bar{\carr_0},\Bar{\carr_1})
\left( \acarr_0 \atop \acarr_1 \right)\cr
&&\hfill={}& \Re (\Bar{\carr_0}\acarr_0 + \Bar{\carr_1}\acarr_1).
\cr}}\label{eq:o8 herm form}
\end{equation}
The designation ``hermitian'' is somewhat misleading, since the octonionic
representation $\rep_{8,0}$ is Majorana, i.e., essentially real, which is also
the reason for taking the real part above.  So the spinor adjoint is given by
\begin{equation}
\adj\carr := \Sphf(\carr) = \carr^\dagger\Sphf = \carr^\dagger
\qquad (\carr\in \Carr).
\end{equation}
Apart from the scalar, we form tensors as spinor bilinears as in
(\ref{eq:tensor bilinear}):
\begin{equation}
Y_{k_1\ldots k_r} := \Re \adj\carr \Rep_{k_1}\ldots\Rep_{k_r} \acarr.
\end{equation}
Since the real part of an associator vanishes (\ref{eq:Re comm=0}) and $\Sphf$
is real, we may associate the matrices sandwiched between the two spinors
differently:
\begin{equation}\acsz\begin{array}{rl}
\Re \adj\carr \Rep_{k_1}\ldots\Rep_{k_r} \acarr
={}& \Re (\carr^\dagger \Sphf)
\bigl[\Rep_{k_1}(\ldots(\Rep_{k_r} \acarr)\ldots)\bigr]
\\ \noalign{\smallskip}
={}& \Re \bigl[(\carr^\dagger \Sphf)
\Rep_{k_1}\bigr](\Rep_{k_2}(\ldots(\Rep_{k_r} \acarr)\ldots))
\\ \noalign{\smallskip}
={}& \Re \bigl[\carr^\dagger (\Sphf \Rep_{k_1})\bigr]
(\Rep_{k_2}(\ldots(\Rep_{k_r} \acarr)\ldots))
\\ \noalign{\smallskip}
={}& \Re \bigl[(\carr^\dagger \Rep_{k_1}^\dagger)\Sphf \bigr]
(\Rep_{k_2}(\ldots(\Rep_{k_r} \acarr)\ldots))
\\ \noalign{\smallskip}
={}& \Re \adj{\Rep_{k_1}\carr}(\Rep_{k_2}(\ldots(\Rep_{k_r}
\acarr)\ldots)).
\end{array}\end{equation}
Since the real part of a commutator vanishes also, we may cyclicly
permute, if a trace is included
\begin{equation}\acsz\begin{array}{rl}
\Re \adj\carr \Rep_{k_1}\ldots\Rep_{k_r} \acarr
={}& \Re \tr{\adj\carr (\Rep_{k_1}(\ldots(\Rep_{k_r} \acarr)\ldots))}
= \Re \tr{(\Rep_{k_1}(\ldots(\Rep_{k_r} \acarr)\ldots))\adj\carr}
\\ \noalign{\smallskip}
={}&\Re\tr{
(\Rep_{k_2}(\ldots(\Rep_{k_r} \acarr)\ldots)\adj\carr)\Rep_{k_1}}.
\end{array}\end{equation}
For the vector covariant, we have a particular expression
\begin{equation}\acsz\begin{array}{rl}
y_k := \Re \adj\carr \Rep_k\acarr
={}& \Re (\Bar{\carr_0},\Bar{\carr_1})
\pmatrix{0&\iu_k\cr\Bar{\iu_k}&0\cr}
\pmatrix{\acarr_0\cr\acarr_1\cr}
= \Re (\Bar{\carr_0}\iu_k\acarr_1
+ \Bar{\carr_1}\Bar{\iu_k}\acarr_0)
\\ \noalign{\smallskip}
={}& \Re (\iu_k\acarr_1\Bar{\carr_0}
+ \acarr_0\Bar{\carr_1}\Bar{\iu_k})
= \Re (\carr_0\Bar{\acarr_1}\Bar{\iu_k}
+ \acarr_0\Bar{\carr_1}\Bar{\iu_k})
\\ \noalign{\smallskip}
={}& (\carr_0\Bar{\acarr_1} + \acarr_0\Bar{\carr_1})_k\,,
\end{array}\end{equation}
where we used once for part of the expression that the real part does not
change under octonionic conjugation.  So we can express the $k$-th component
of $y$ by the $k$-th component of an octonionic product, which allows us to
write $\Slash{y}$ without the use of the matrix representations of the basis
elements:
\begin{equation}\acsz\begin{array}{rl}
\Slash{y} ={}& \pmatrix{0&y\cr\Bar{y}&0\cr}
= \Rep^k \Re \adj\carr \Rep_k\acarr
\\ \noalign{\smallskip}
={}& \pmatrix{0&\carr_0\Bar{\acarr_1} + \acarr_0\Bar{\carr_1}\cr
\Bar{(\carr_0\Bar{\acarr_1} + \acarr_0\Bar{\carr_1})}&0\cr}.
\label{eq:vector covariant}
\end{array}\end{equation}

\subsection{Orthogonal transformations}\label{subsec:8D trafos}

{}From section~\ref{subsec:Cg} we know the action of the Clifford group on
vectors (\ref{eq:tensor action}) and spinors (\ref{eq:spinor action}).  The
condition (\ref{eq:Cg normalized}) shows how to divide out $\R^*$ to obtain
the orthogonal group.  So elements of $V$ satisfying
\begin{equation}
\mant(u)\vm u = 1 \qquad \iff \qquad u \vm u = g(u,u) = \normsq{u} = 1
\end{equation}
generate the orthogonal transformations via
\begin{eqnarray}
& \Sslash{x^\prime}{4} =
& (\rep\circ\Cac_u)(x) = \Slash{u}\Slash{x}\Slash{u}
= \pmatrix{0 & u\Bar{x}u\cr\Bar{u}x\Bar{u} & 0\cr},
\label{eq:vector O(8)}\\ \noalign{\smallskip}
& \carr^\prime =
& \Csac_u(\carr)
= \Slash{u}\carr = \pmatrix{u\carr_1\cr\Bar{u}\carr_0\cr}.
\label{eq:spinor O(8)}
\end{eqnarray}
The Moufang (\ref{eq:Moufang}) identities ensure that (\ref{eq:vector O(8)})
is unambiguous and even holds under the action of left multiplication, which
can be seen in the example, $(x\vm\carr)^\prime = x^\prime\vm\carr^\prime$:
\begin{equation}\acsz\begin{array}{rl}
\Sslash{x^\prime}{4}\carr^\prime
={}& \pmatrix{0 & u\Bar{x}u\cr \Bar{u}x\Bar{u} & 0\cr}
\pmatrix{u\carr_1\cr\Bar{u}\carr_0\cr}
= \pmatrix{(\Bar{u}x\Bar{u})(u\carr_1)\cr (u\Bar{x}u)(\Bar{u}\carr_0)\cr}
\\ \noalign{\smallskip}
={}& \pmatrix{\Bar{u}(x(\Bar{u}(u\carr_1)))\cr
u(\Bar{x}(u(\Bar{u}\carr_0)))\cr}
= \pmatrix{\Bar{u}(x((\Bar{u}u)\carr_1))\cr u(\Bar{x}((u\Bar{u})\carr_0))\cr}
\\ \noalign{\smallskip}
={}& \normsq{u} \pmatrix{\Bar{u}(x\carr_1)\cr u(\Bar{x}\carr_0)\cr}
= \Slash{u}(\Slash{x}\carr)
\\ \noalign{\smallskip}
={}& (\Slash{x}\carr)^\prime.
\end{array}\end{equation}
The third Moufang identity guarantees that the vector covariant (\ref{eq:vector
covariant}) of two spinors transform correctly:
\begin{equation}\acsz\begin{array}{rl}
\Sslash{y^\prime}{4} ={}& \pmatrix{0&u\Bar{y}u\cr \Bar{u}y\Bar{u}&0\cr}
= \pmatrix{0&u\Bar{(\carr_0\Bar{\acarr_1}
+ \acarr_0\Bar{\carr_1})}u\cr
\Bar{u}(\carr_0\Bar{\acarr_1} + \acarr_0\Bar{\carr_1})\Bar{u}&0\cr}
\\ \noalign{\smallskip}
={}& \pmatrix{0&\Bar{(\Bar{u}(\carr_0\Bar{\acarr_1}
+ \acarr_0\Bar{\carr_1})\Bar{u})}\cr
\Bar{u}(\carr_0\Bar{\acarr_1} + \acarr_0\Bar{\carr_1})\Bar{u}&0\cr}
\\ \noalign{\smallskip}
={}& \pmatrix{0&\Bar{[(\Bar{u}\carr_0)(\Bar{\acarr_1}\Bar{u})
+ (\Bar{u}\acarr_0)(\Bar{\carr_1}\Bar{u})]}\cr
(\Bar{u}\carr_0)(\Bar{\acarr_1}\Bar{u})
+ (\Bar{u}\acarr_0)(\Bar{\carr_1}\Bar{u})&0\cr}
\\ \noalign{\smallskip}
={}& \pmatrix{0&\Bar{[(\Bar{u}\carr_0)\Bar{(u\acarr_1)}
+ (\Bar{u}\acarr_0)\Bar{(u\carr_1)}]}\cr
(\Bar{u}\carr_0)\Bar{(u\acarr_1)}
+ (\Bar{u}\acarr_0)\Bar{(u\carr_1)}&0\cr}
\\ \noalign{\smallskip}
={}& \Rep^k \Re \adj{\carr^\prime} \Rep_k\acarr^\prime.
\end{array}\end{equation}

According to (\ref{eq:one vector fixed}), simple orthogonal transformations
are generated by pairs $(u,v) \in V \times V$, where we take $v = \ob_0$
fixed and $\normsq{u} = 1$:
\begin{eqnarray}
& \Sslash{x^\prime}{4} =
& (\rep_{8,0}\circ\Cac_{(u\vm v)})(x)
= \Slash{u}\Slash{1}\Slash{x}\Slash{1}\Slash{u}
= \pmatrix{0 & uxu\cr\Bar{u}\Bar{x}\Bar{u} & 0\cr}
= \pmatrix{0 & uxu\cr\Bar{(uxu)} & 0\cr},
\label{eq:vector SO(8)}\\ \noalign{\smallskip}
& \carr^\prime =
& \Csac_{(u\vm v)}(\carr)
= \Slash{u}\Slash{1}\carr = \pmatrix{u\carr_0\cr\Bar{u}\carr_1\cr}.
\label{eq:spinor SO(8)}
\end{eqnarray}
Choosing the fixed vector to be $\ob_0$ allows significant simplification,
since its representation $\Rep_0$ is real.  How to construct any orthogonal
transformation from these generators is thoroughly explained in
\cite{Cl:Lorentz}.
These transformation properties imply that the definition of the spinor
covariants in section~\ref{subsec:herm conj rep} is consistent.  For example,
\begin{equation}
\Re \adj{\carr^\prime}\Sslash{x^\prime}{4}\acarr^\prime
= \Re \adj{\Slash{u}\Slash{1}\carr}\,
\Slash{u}\Slash{1}\Slash{x}\Slash{1}\Slash{u}
\,\Slash{u}\Slash{1}\acarr
= \Re \carr^\dagger\Slash{u}^\dagger\Slash{1}^\dagger \Sphf
\Slash{u}\Slash{1}\,\Slash{x}\,\acarr
= \Re \carr^\dagger\Sphf\Slash{1}\Slash{u}\Slash{u}\Slash{1}\,\Slash{x}\,\acarr
= \Re \adj\carr\,\Slash{x}\,\acarr\,.
\label{eq:scalar}\end{equation}

\subsection{Related representations using the opposite octonionic algebra
$\O_\opp$}

As pointed out in section~\ref{subsec:herm conj rep}, transposition and
octonionic conjugation are not (anti-)automorphisms of octonionic matrix
multiplication.  However, we can find (anti-)isomorphisms to matrix algebras
by using the opposite octonionic algebra $\O_\opp$.  We define the octonionic
conjugate representation $\Bar{\rep}$ of an octonionic representation
$\rep:\alg \to \mat_\dmW(\O)$ by
\begin{equation}
\Bar{\rep}: \vtop{\halign{$\hfil#$&$\hfil{}#{}\hfil$&$#\hfil$\cr
\alg & \to & \mat_\dmW(\Bar{\O_\opp})\cr
\noalign{\smallskip}
a & \mapsto & \Bar{\rep}(a):=\Bar{(\rep(a))_\opp}.
\cr}}
\end{equation}
Octonionic products are now to be evaluated in the opposite algebra as  is
indicated in the following examples.  First we consider the action of
$\Bar{\rep_{8,0}}(x)$ for $x \in V$ on an element
$\Bar{\carr}$ of the carrier space $\Bar{\Carr_\opp}$
\begin{equation}\acsz\begin{array}{rl}
\Bar{\rep_{8,0}}(x)\Bar{\carr_\opp}
={}& \left( \Bar{\pmatrix{0&x\cr\Bar{x}&0\cr}}
\Bar{\pmatrix{\carr_0\cr\carr_1\cr}} \right)_\opp
= \left( \pmatrix{0&\Bar{x}\cr x&0\cr}
\pmatrix{\Bar{\carr_0}\cr\Bar{\carr_1}\cr} \right)_\opp
\\ \noalign{\smallskip}
={}& \pmatrix{\Bar{\carr_1}\Bar{x}\cr\Bar{\carr_0}x\cr}
\\ \noalign{\smallskip}
= \Bar{(\rep_{8,0}(x)\carr)}
={}& \Bar{\pmatrix{x\carr_1\cr \Bar{x}\carr_0\cr}}
= \pmatrix{\Bar{(x\carr_1)}\cr\Bar{(\Bar{x}\carr_0)}\cr}
\end{array}\end{equation}
So in this representation the action on the carrier space is effectively
right multiplication by octonions.

We check that $\Bar{\rep_{8,0}}$ is indeed a representation.  Let $u,v
\in V$, then
\begin{equation}\acsz\begin{array}{rl}
\Bar{\rep_{8,0}}(u) \Bar{\rep_{8,0}}(v)
={}&\Bar{\pmatrix{0&u\cr\Bar{u}&0\cr}_\opp}
\Bar{\pmatrix{0&v\cr\Bar{v}&0\cr}_\opp}
= \left(\pmatrix{0&\Bar{u}\cr u&0\cr}
\pmatrix{0&\Bar{v}\cr v&0\cr}\right)_\opp
\\ \noalign{\smallskip}
={}& \pmatrix{v\Bar{u}&0\cr 0&\Bar{v}u\cr}
\\ \noalign{\smallskip}
= \Bar{\rep_{8,0}}(u \vm v)
={}& \Bar{\left(\rep_{8,0}(u \vm v)\right)}
= \Bar{\left(\Slash{u}\Slash{v}\right)}
\\ \noalign{\smallskip}
={}& \Bar{\pmatrix{u\Bar{v}&0\cr 0&\Bar{u}v\cr}}.
\end{array}\end{equation}
In both cases the subscript ``opp'' indicates that the remaining products are
to be done in the opposite octonionic algebra.  However the final result is to
be interpreted as an element of $\Bar{\Carr_\opp}$
(resp.\ $\mat_2(\Bar{\O_\opp})$).

Since
\begin{equation}
\Rep_0 \Rep_k = \Bar{(\Rep_k)} \Rep_0 \qquad (0\leq k \leq 7),
\end{equation}
we define the map
\begin{equation}
\Cc: \vtop{\halign{$\hfil#$&$\hfil{}#{}\hfil$&$#\hfil$\cr
\Carr & \to & \Bar{\Carr_\opp}\cr
\noalign{\smallskip}
w & \mapsto & \Cc(w) := \Rep_0 w_\opp.
\cr}}
\end{equation}
This map gives rise to an operation on $\Carr$ which is analogous to charge
conjugation:
\begin{equation}
\carr_\Cc := \Bar{C(w)} = \Rep_0 \Bar\carr
= \pmatrix{\Bar{\carr_1}\cr\Bar{\carr_0}\cr} \;\in \Carr.
\end{equation}
Let us examine how $\carr_\Cc$ transforms under an orthogonal transformation:
\begin{equation}\acsz\begin{array}{rl}
(\carr_\Cc)^\prime ={}& \Slash{u}\carr_\Cc
= \pmatrix{0 & u\cr\Bar{u} & 0\cr} \pmatrix{\Bar{\carr_1}\cr\Bar{\carr_0}\cr}
\\ \noalign{\smallskip}
={}&{} \pmatrix{u\Bar{\carr_0}\cr\Bar{u}\Bar{\carr_1}\cr}
= \Bar{\pmatrix{\carr_0\Bar{u}\cr\carr_1 u\cr}}
\\ \noalign{\smallskip}
={}& \Bar{\pmatrix{\Bar{u}\carr_0\cr u\carr_1\cr}_\opp}
= \Bar{\left(\left[
\pmatrix{0 & \Bar{u}\cr u & 0\cr}\pmatrix{0 & 1\cr 1 & 0\cr}
\pmatrix{\carr_0\cr \carr_1\cr}\right]_\opp\right)}
\\ \noalign{\smallskip}
={}& \Bar{\left(\left[\Bar{\rep_{8,0}}(u)\Rep_0 \carr\right]_\opp\right)}
\\ \noalign{\smallskip}
={}& \Rep_0\Bar{\left(\left[\rep_{8,0}(u) \carr\right]_\opp\right)}.
\end{array}
\label{eq:Cc spinor}
\end{equation}
So the map $\Cc$ almost intertwines $\Bar{\rep_{8,0}}$ and $\rep_{8,0}$,
except that the opposite octonionic algebra has to be included explicitly:
\begin{equation}
(\carr_\Cc)^\prime \not= \Bar{(\Cc(\carr^\prime))}
= \Bar{\left[\Rep_0\pmatrix{u\carr_1\cr \Bar{u}\carr_0\cr}\right]}
= \Bar{\pmatrix{\Bar{u}\carr_0\cr u\carr_1\cr}}.
\label{eq:Cc problem}\end{equation}
Of course, octonionic conjugation intertwines $\Bar{\rep_{8,0}}$ and
$\rep_{8,0}$, but there is no octonionic linear transformation that does the
job.

Related to matrix transposition we obtain another representation $\Ch\rep$
involving $\O_\opp$:
\begin{equation}
\Ch\rep: \vtop{\halign{$\hfil#$&$\hfil{}#{}\hfil$&$#\hfil$\cr
\alg & \to & \T\mat_\dmW(\O_\opp)\cr
\noalign{\smallskip}
a& \mapsto & \Ch\rep(a):=\T{(\rep(\mant(a)))_\opp}
:\vtop{\halign{$\hfil#$&$\hfil{}#{}\hfil$&$#\hfil$\cr
\T\Carr_\opp & \to & \T\Carr_\opp\cr
\noalign{\smallskip}
\T\carr & \mapsto &
\Ch\rep(a)(\T\carr) = \left(\T\carr\rep(\mant(a))\right)
= \T{\left(\T\rep(\mant(a))\carr\right)}_\opp
\,.\cr}}
\cr}}
\end{equation}
The verification of $\Ch\rep(a \vm b) = \Ch\rep(a) \Ch\rep(b)$ is another
exercise in applying opposite algebras:
\begin{equation}\acsz\begin{array}{rl}
\Ch\rep(a \vm b)
={}& \T{\left(\rep(\mant(a \vm b)) \right)}
= \T{\left(\rep(\mant(b) \vm \mant(a)) \right)}
\\ \noalign{\smallskip}
={}& \T{\left(\rep(\mant(b)) \rep(\mant(a)) \right)}
= \left(\T{\left(\rep(\mant(a)) \right)}
\T{\left(\rep(\mant(b))\right)}\right)_\opp
\\ \noalign{\smallskip}
={}& \Ch\rep(a) \Ch\rep(b).
\end{array}\end{equation}
The map that almost intertwines $\Ch\rep_{8,0}$ and $\rep_{8,0}$ is
\begin{equation}
\Spbl: \vtop{\halign{$\hfil#$&$\hfil{}#{}\hfil$&$#\hfil$\cr
\Carr & \to & \T\Carr_\opp\cr
\noalign{\smallskip}
w & \mapsto & \Spbl(w) := \T{w_\opp} \Rep_0,
\cr}}
\end{equation}
since
\begin{equation}
\Rep_0 \Rep_k = \T{(\Rep_k)} \Rep_0 \qquad (0\leq k \leq 7).
\end{equation}

We have seen that the non-commutativity of the octonions has important
consequences for representations that are related by octonionic conjugation
and matrix transposition.  The natural space for these representations to act
on involves the opposite octonionic algebra, which prevents us from finding
intertwining maps.  Therefore special care should be taken when octonionic
conjugation or matrix transposition is part of a manipulation involving
octonionic spinors.
However, this additional freedom of choosing different multiplication rules
for different representations and carrier spaces may turn out to be
advantageous in applications.  In the following section we will observe how
more general changes of multiplication rules further increase the flexibility
of an octonionic representation.

\subsection{Octonionic spinors as elements of minimal left ideals}
\label{subsec:twisted spinors}

In this section we take a different perspective on octonionic spinors,
regarding them as elements of a minimal left ideal which is generated by a
certain primitive idempotent.  The choice of an idempotent will turn out to be
equivalent to the choice of a basis of the carrier space of the
representation, which may be understood as a change of the multiplication rule
of the octonions.  An immediate application of the ideas presented here can be
found in \cite{Cl:superparticle}.

In a real or complex representation $\rep : \alg \to \End_\F(\Carr,\Carr)$ of
dimension~$\dmW$ an idempotent is given by an $\dmW\times\dmW$-matrix $\prmid$
satisfying the minimal polynomial $\prmid(\prmid -1) = 0$.  Therefore,
$\prmid$ can be diagonalized with eigenvalues $0$ and $1$.  If the
representation is onto and the idempotent is primitive, then  $\prmid$ is of
rank $1$ and there is a transformation such that $\prmid$ takes the form
\begin{equation}
\prmid = \pmatrix{ 1 & 0 &\ldots & 0\cr
0 & 0 & \ldots&  0 \cr
\vdots & \vdots & \ddots & \vdots \cr
0 & 0 & \ldots&  0 \cr}
= \pmatrix{ 1\cr 0 \cr \vdots \cr 0 \cr}
\matrix{ \pmatrix{ 1& 0 & \ldots & 0 \cr} \cr  \phantom{\vdots}\cr  \cr \cr}.
\end{equation}
So for a surjective representation a primitive idempotent is represented by a
matrix of the form
\begin{equation}
\prmid = \gen \,\T\gene, \quad \T\gene\gen =1 \qquad (\gen,\gene \in \Carr).
\end{equation}
The minimal left ideal $\Id = \alg \vm \prmid$ generated by $\prmid$ in this
representation consists of matrices with linearly dependent columns.
Therefore, the action of the Clifford algebra on the minimal left ideal $\Id$
is determined by $\gen$ alone.  So the relevant choices of primitive
idempotents are given by the choices for $\gen$.  The choice of a basis for
$\Id$ is still arbitrary at this point.  For the octonionic case, however,
there is a connection between the choice of $\gen$ and a multiplication rule.

In terms of the octonionic representation $\rep_{8,0}$ we have
$\gen = \left({\gen_0\atop\gen_1}\right)\in \O^2$.
For $\gen$ to correspond to an even primitive idempotent $\prmid$, one of its
components has to vanish.  (Note that even elements of the Clifford algebra
are represented by diagonal matrices, whereas for odd elements the matrices
have vanishing diagonal components.)  We may also normalize $\gen$.  So let
$\gen =\left({\rho\atop 0}\right)$ with $\normsq{\rho} =1$.
(A vanishing upper component leads to similar results.)  A natural
parametrization of the spinor space $\Id$ is given by
\begin{equation}
\sp := \left(\Sslash{\sp_1}{4} + \Sslash{\sp_0}{4} \Rep_0\right) \gen
= \pmatrix{ \sp_0&  \sp_1\cr\Bar{\sp_1} & \Bar{\sp_0}\cr}
\pmatrix{ \rho \cr  0 \cr}
= \pmatrix{ \sp_0\rho \cr  \Bar{\sp_1}\rho \cr}.
\end{equation}
We interpret the octonions $\sp_0$ and $\sp_1$ as the new labels or components
for the spinor $\sp$.  The choice of this parametrization is natural, since it
is up to octonionic conjugation the only one that involves only one left
multiplication by an octonion.  How does the Clifford
algebra act in terms of the new spinor components?  For $x \in V$
\begin{equation}\acsz\begin{array}{rl}
\sp^\prime ={}& \Slash{x} \sp
= \pmatrix{  x(\Bar{\sp_1}\rho) \cr \Bar{x}(\sp_0\rho) \cr }
= \pmatrix{ \sp_0^\prime\rho \cr  \Bar{\sp_1}^\prime\rho \cr}
\\ \noalign{\smallskip}
\implies \quad \sp_0^\prime ={}& [x(\Bar{\sp_1}\rho)]\Bar\rho,
\quad \sp_1^\prime = \rho[(\Bar\rho\Bar{\sp_0})x],
\end{array}\end{equation}
which leads to two other versions of the ``$X$-product'' (\ref{eq:x-product})
with $X=\rho$:
\begin{equation}\acsz\begin{array}{rl}
\sp_0^\prime ={}& [x(\Bar{\sp_1}\rho)]\Bar\rho
= [(x\rho\Bar\rho)(\Bar{\sp_1}\rho)]\Bar\rho
= [((x\rho)\Bar\rho)(\Bar{\sp_1}\rho)]\Bar\rho
= (x\rho)[\Bar\rho(\Bar{\sp_1}\rho)\Bar\rho],
\\ \noalign{\smallskip}
={}& (x\rho)(\Bar\rho\Bar{\sp_1}) = x \cpr_\rho \Bar{\sp_1},
\\ \noalign{\medskip}
\sp_1^\prime ={}& \rho[(\Bar\rho\Bar{\sp_0})x]
=  \rho[(\Bar\rho\Bar{\sp_0})(\rho\Bar\rho x)]
=  \rho[(\Bar\rho\Bar{\sp_0})(\rho(\Bar\rho x))]
=  [\rho(\Bar\rho\Bar{\sp_0})\rho](\Bar\rho x),
\\ \noalign{\smallskip}
={}&  (\Bar{\sp_0}\rho)(\Bar\rho x)
= \Bar{\sp_0} \cpr_\rho x,
\end{array}\end{equation}
where the fourth equality uses (\ref{eq:Moufang}).  Octonionic conjugation is
also an antiautomorphism of the ``$X$-product'', which gives the
transformation behavior of $\Bar{\sp_1}$ as new spinor component.
\begin{equation}\acsz\begin{array}{rl}
\Bar{{\sp_1^\prime}} = {\Bar{\sp_1}}^\prime
={}& \Bar{(\Bar{\sp_0} \cpr_\rho x)} = \Bar{x} \cpr_\rho \sp_0,
\\ \noalign{\medskip}
\pmatrix{ \sp_0^\prime\cr  \Bar{{\sp_1^\prime}}\cr}
={}& \Slash{x} \cpr_\rho \pmatrix{ \sp_0 \cr\Bar{\sp_1}\cr}.
\end{array}\end{equation}
Therefore, choosing $\sp_0$ and $\Bar{\sp_1}$ as new spinor components is
equivalent to replacing the original octonionic product with the
``$\rho$-product''.  We confirm this result for the scalar formed out of two
spinors (compare (\ref{eq:o8 herm form})):
\begin{equation}\acsz\begin{array}{rl}
\Re \adj\sp \sp^\prime
={}& \Re \matrix{\pmatrix{ \Bar\rho \Bar{\sp_0} &\Bar\rho \sp_1 \cr}\cr\cr}
\pmatrix{ \sp_0^\prime\rho \cr  \Bar{{\sp_1^\prime}}\rho \cr}
= \Re [(\Bar\rho \Bar{\sp_0})(\sp_0^\prime\rho)
+ (\Bar\rho \sp_1)(\Bar{{\sp_1^\prime}}\rho)]
\\ \noalign{\smallskip}
={}& \Re [(\sp_0^\prime\rho)(\Bar\rho \Bar{\sp_0})
+ (\Bar{{\sp_1^\prime}}\rho)(\Bar\rho \sp_1)]
= \Re (\sp_0^\prime \cpr_\rho \Bar{\sp_0}
+ \Bar{{\sp_1^\prime}} \cpr_\rho \sp_1)
\\ \noalign{\smallskip}
={}& \Re (\sp_0^\prime \Bar{\sp_0} + \Bar{{\sp_1^\prime}} \sp_1)
= \Re (\Bar{\sp_0} \cpr_\rho \sp_0^\prime
+ \sp_1 \cpr_\rho \Bar{{\sp_1^\prime}})
\end{array}\end{equation}
as well as the vector (compare (\ref{eq:vector covariant}))
\begin{equation}\acsz\begin{array}{rl}
\Rep^k \Re \adj\sp \Rep_k\sp^\prime
={}& \pmatrix{0&
(\sp_0\rho)(\Bar{\rho}\sp^\prime_1) + (\sp^\prime_0\rho)(\Bar{\rho}\sp_1)\cr
\Bar{[
(\sp_0\rho)(\Bar{\rho}\sp^\prime_1) + (\sp^\prime_0\rho)(\Bar{\rho}\sp_1)]}
&0\cr}
\\ \noalign{\smallskip}
={}& \pmatrix{0&
\sp_0\cpr_\rho\sp^\prime_1 + \sp^\prime_0\cpr_\rho\sp_1\cr
\Bar{[\sp_0\cpr_\rho\sp^\prime_1 + \sp^\prime_0\cpr_\rho\sp_1]}
&0\cr}.
\end{array}\end{equation}

Of course, orthogonal transformations, as described in section~\ref{subsec:8D
trafos}, also induce a change of basis on the spinor space.  The corresponding
change of the octonionic multiplication rule is more complex since the real
part is no longer fixed (compare section~\ref{subsec:mult tables}).

\section{Other octonionic representations}\label{sec:other reps}

In this section, we point out the constructions of octonionic representations
related to $\rep_{8,0}$.  We follow the program outlined in section
\ref{subsec:reps}.
First we shrink the representation of $\Cl(8,0)$ to obtain one of $\Cl_0(8,0)
\isom \Cl(0,7)$ and further of $\Cl(0,6)$.  Then we look at the extension to a
representation of $\Cl(9,1)$, which is of particular importance, since it
applies to superstring and superparticle models.

\subsection{$\Cl_0(8,0)$ and $\Cl(0,7)$}

Restricting the representation $\rep_{8,0}$ to $\Cl_0(8,0) \isom \Cl_0(0,8)$
produces a faithful representation with the generators
\begin{equation}
\Rep_0\Rep_k = \rep_{8,0}(\ob_0\vm\ob_k)
= \pmatrix{\iu_k & 0\cr 0&\Bar{\iu_k}\cr}
= \pmatrix{\iu_k & 0\cr 0& -\iu_k\cr} \qquad (1\leq k \leq 7).
\label{eq:faithful 0,7}
\end{equation}
So $\Cl_0(8,0)$ is represented by diagonal matrices, i.e., this
representation decomposes into two irreducible representations given by the
two elements on the diagonal.  By the isomorphism $\Cl_0(8,0) \isom \Cl(0,7)$
(\ref{eq:even subalgebra}), these two are also irreducible representations
$\rep_{0,7}^\pm:\Cl(0,7)\to\mat_1(\O)=\O$,
\begin{eqnarray}
\rep_{0,7}^\pm(\ob_k)&:=&\pm\iu_k \qquad (1\leq k \leq 7)\\
\iff \qquad
\rep_{0,7}^\pm(x)&:=&\pm x = \pm \Im x\qquad (x \in V = \R^7).
\label{eq:7orep}
\end{eqnarray}
So we identify $V=\R^7$ with the purely imaginary subspace of the octonions
$\Im\O$.
A faithful representation of $\Cl(0,7)$ is found by letting $\rep_{0,7}(\ob_k)
= \Rep_0\Rep_k$ in (\ref{eq:faithful 0,7}):
\begin{equation}
\rep_{0,7}:= \rep_{0,7}^+ \oplus \rep_{0,7}^-
\quad\iff\quad \rep_{0,7}(a)
= \pmatrix{\rep_{0,7}^+(a)&0\cr 0&\rep_{0,7}^-(a)\cr}.
\end{equation}
A hermitian form $\Sphf^\prime: \O^\pm \to \O^{\pm\,\dagger}$ on the carrier
space of an irreducible representation is given by
\begin{equation}
\Sphf^\prime(\carr) := \Bar\carr
\end{equation}
with the property
\begin{equation}
\Sphf^\prime \rep_{0,7}^\pm(\ob_k)
= - \rep_{0,7}^{\pm\,\dagger}(\ob_k) \Sphf^\prime
= - \Bar{\left(\rep_{0,7}^\pm(\ob_k)\right)}
\qquad (1\leq k\leq 7).
\end{equation}
Thus the form $\Sphf^\prime$ intertwines $\rep_{0,7}^\pm$ and
$\rep_{0,7}^{\pm\,\dagger}\circ\maut\circ\mant$:
\begin{equation}
\Sphf^\prime \circ \rep_{0,7}^\pm(a)
= \left(\rep_{0,7}^\pm((\maut\circ\mant)(a))\right)^\dagger \circ \Sphf^\prime
\qquad (a \in\Cl(0,7)).
\end{equation}
There is no sesquilinear form satisfying (\ref{eq:sesquilinear form}) on a
carrier space of the irreducible representation.  However, one can intertwine
$\rep_{0,7}^+$ and $\rep_{0,7}^-$ to obtain such a form on the carrier space
$2\O = \O^+ \oplus \O^-$ of the faithful representation that swaps the two
copies $\O^+$ and $\O^-$ of $\O$ since
\begin{equation}
\rep_{0,7}^\pm(a) = \Bar{\left(\rep_{0,7}^\mp(\mant(a))\right)}
\qquad (a \in\Cl(0,7)).
\end{equation}
$\Sphf$, defined by
\begin{equation}
\Sphf(\carr^+ \oplus \carr^-) := \Bar{{\carr^-}}\oplus\Bar{{\carr^+}}
= \adj\carr
\quad\iff\quad
\Sphf := \pmatrix{0 & 1\cr 1 & 0\cr} = \sig1,
\end{equation}
satisfies
\begin{equation}\acsz\begin{array}{rc}
&\Sphf \circ \rep_{0,7}(a)
= \rep_{0,7}^\dagger(\mant(a)) \circ \Sphf \qquad (a \in \Cl(0,7))
\\ \noalign{\smallskip}
\iff\quad
&\Sphf \rep_{0,7}(\ob_k) = \rep_{0,7}^\dagger(\ob_k) \Sphf
\qquad (1\leq k\leq 7).
\end{array}\end{equation}
Simple orthogonal transformations are generated by unit vectors $u \in \Im\O$,
$\normsq{u} = -u^2 = 1$ via
\begin{eqnarray}
& x^\prime =
& (\rep_{0,7}^\pm\circ\Cac_u)(x) = (\pm u)x(\pm u)^{-1} = ux\Bar{u} = -uxu,
\label{eq:vector SO(7)}\\ \noalign{\smallskip}
& \carr^{\pm\,\prime} =
& \Csac_u(\carr)
= \pm u\carr.
\label{eq:spinor SO(7)}
\end{eqnarray}
Since the real part of $u$ vanishes, $u^{-1} = -u$.  Therefore, the
transformations have the same form as (\ref{eq:vector O(8)}) and
(\ref{eq:spinor O(8)}) up to signs and the Moufang identities ensure the
compatibility of the spinor and vector transformations as before.  As is seen
from (\ref{eq:hom Cg O odd}), improper rotations, for example, inversion of
$\R^7$, $x \mapsto -x = \Bar{x}$, is not described by the action of the
Clifford group for odd $\dmV$.  In fact, inversion is equivalent to octonionic
conjugation or switching from $\rep_{0,7}^\pm$ to $\rep_{0,7}^\mp$.  In order
to implement inversion we need to use the faithful representation:
\begin{equation}\acsz\begin{array}{rl}
\Sslash{x^\prime}{4} ={}& -\sig2\Slash{x}\sig2
= -\pmatrix{0 & -1\cr 1 & \phm0\cr}\pmatrix{x & \phm0\cr 0 & -x\cr}
\pmatrix{0 & -1\cr 1 & \phm0\cr} = -\Slash{x},
\label{eq:vector O(7)}\\ \noalign{\smallskip}
\carr^\prime = \pmatrix{\carr^{+\,\prime}\cr \carr^{-\,\prime}\cr} ={}&
\sig2\carr^\prime
= \pmatrix{0 & -1\cr 1 & \phm0\cr} \pmatrix{\carr^+\cr\carr^-\cr}
= \pmatrix{\phm\carr^-\cr -\carr^+\cr}.
\label{eq:spinor O(7)}
\end{array}\end{equation}
The transformation preserves scalars:
\begin{equation}
\adj\carr \Slash{x} \acarr
= \carr^\dagger \sig1 (-\sig2)(-\sig2) \Slash{x} \sig2 \sig2 \acarr
= (\carr^\dagger \sig2 \sig1) [-\sig2 \Slash{x} \sig2] (\sig2 \acarr)
= \adj\carr^\prime \Sslash{x^\prime}{4} \acarr^\prime.
\end{equation}

\subsection{$\Cl_0(0,7)$ and $\Cl(0,6)$}

Shrinking a representation of $\Cl(0,7)$ further leads to the smallest
Clifford algebra that has the octonions as a natural carrier space for a
representation. Both irreducible representations $\rep_{0,7}^+$ and
$\rep_{0,7}^-$ agree on the even Clifford algebra $\Cl_0(0,7)\isom\Cl_0(7,0)$.
Their restriction is an irreducible representation given by the generators
\begin{equation}
\rep_{0,7}^\pm(\ob_k \vm \ob_7) = \iu_k \iu_7 \qquad (1\leq k\leq 6),
\end{equation}
which act by successive left multiplication on the carrier space $\Carr = \O$.
Again by the isomorphism $\Cl_0(0,7) \isom \Cl(0,6)$ (\ref{eq:even
subalgebra}), we obtain a faithful and irreducible representation of
$\Cl(0,6)$, $\rep_{0,6} : \Cl(0,6) \to \mat_1(\O) = \O$,
\begin{eqnarray}
\rep_{0,6}(\ob_k) &:=& \iu_k \iu_7 \qquad (1\leq k\leq 6),
\\ \iff \qquad
\rep_{0,6}^\pm(x)&:=&x \iu_7 \qquad (x \in V = \R^6).
\label{eq:6orep}
\end{eqnarray}
$V=\R^6$ is identified with the imaginary subspace of $\O$ with vanishing
7-component, $\{x\in\Im\O: x^7=0\}$.
The volume form $\vol$ is represented by
\begin{equation}
\rep_{0,6}(\vol) = \rep_{0,6}(\ob_1 \vm \ob_2 \vm \cdots \vm \ob_6)
= \iu_1 \iu_7 \iu_2 \iu_7 \ldots \iu_6 \iu_7
= - \iu_1 \iu_2 \ldots \iu_6
= \iu_7,
\end{equation}
according to (\ref{eq:complete product}).
A hermitian form $\Sphf^\prime: \O \to \O^\dagger$ is given by
\begin{equation}
\Sphf^\prime(\carr) := \Bar\carr.
\end{equation}
Orthogonal transformations are generated by unit vectors $u \in \R^6$,
$\normsq{u} = -u^2 = 1$ via
\begin{eqnarray}
& x^\prime =
& (\rep_{0,6}\circ\Cac_u)(x) = (u (\iu_7 x \iu_7) u)
\label{eq:vector O(6)}\\ \noalign{\smallskip}
& \carr^\prime =
& \Csac_u(\carr)
=  u(\iu_7\carr)
\label{eq:spinor O(6)}
\end{eqnarray}
Since these transformations have the same structure as the simple orthogonal
transformations for $V=\R^8$, the Moufang identities ensure their compatibility
and their validity under the interpretation of left multiplication.  Since
$\rep_{0,6}$ is faithful and irreducible and $\Cl(0,6)$ is a $2^6$-dimensional
algebra, we conclude from this section that left multiplication by octonions
generates a 64-dimensional algebra isomorphic to $\mat_8(\R)$.

\subsection{$\Cl(9,1)$}

In this section we will give a little more detail because of the frequent use
of $\Cl(9,1)$ in supersymmetric models.
Starting from $\Cl(8,0)$, we do a Cartan extension (\ref{eq:rep extension}) to
obtain a representation of $\Cl(9,1)$, $\rep_{9,1} : \Cl(9,1)
\to \mat_4(\O)$, given by the generators
\begin{equation}\acsz\begin{array}{rl}
\rep_{9,1}(\ob_k) :={}& \sig1 \otimes \rep_{8,0}(\ob_k)
= \pmatrix{0 & \Rep_k\cr \Rep_k& 0\cr} \quad (0\leq k\leq 7),
\\ \noalign{\smallskip}
\rep_{9,1}(\ob_8) :={}& \sig1 \otimes \rep_{8,0}(\vol)
= \pmatrix{0 & \sig3\cr \sig3 & 0\cr},
\\ \noalign{\smallskip}
\rep_{9,1}(\ob_{-1}) :={}& -\sig2 \otimes \rep_{8,0}(1)
= \pmatrix{\phm 0 & \one\cr -\one & 0\cr},
\end{array}\end{equation}
or equivalently by
\begin{equation}\acsz\begin{array}{rl}
\rep_{9,1}(x) :={}& \Slash{x} = x^\mu\rep_\mu
= \pmatrix{\zero & {\bf X}\cr \tilde{\bf X} & \zero\cr}
\\ \noalign{\medskip\bigskip}
={}& \pmatrix{\zero & \pmatrix{x^+ & x^{\phm} \cr \Bar{x} & x^-\cr}\cr
\pmatrix{-x^- & \phm x^{\phm} \cr \phm\Bar{x} & -x^+\cr} & \zero\cr}\,,
\end{array}\end{equation}
where we defined
\begin{equation}\acsz\begin{array}{rlrl}
{\bf X} :={}& x^\mu\Rep_\mu
= \pmatrix{x^+ & x^{\phm} \cr \Bar{x} & x^-\cr},
\quad&\Rep_8 :={}& \sig3,\; \Rep_{-1} := \one, \;x_\pm:= x_{-1} \pm x_8,
\\ \noalign{\smallskip}
\tilde{\bf X} :={}& x^\mu\tilde\Rep_\mu
= \pmatrix{-x^- & \phm x^{\phm} \cr \phm\Bar{x} & -x^+\cr},
\quad &\tilde\Rep_\mu
:={}& \left\{ {\phm\Rep_\mu, \quad (0\leq \mu \leq 8)
\atop -\Rep_{-1}, \quad (\mu = -1)} \right.\,,
\\ \noalign{\smallskip}
\rep_\mu :={}& \rep_{9,1}(\ob_\mu)
= \pmatrix{\zero & \Rep_\mu\cr \tilde\Rep_\mu & \zero\cr}
\quad \hbox to 0pt{$(-1\leq \mu \leq 8).$\hss}
\end{array}\end{equation}
(Labeling the basis elements of $V = \R^{10}$ by indices ranging from $-1$ to
8, allows us to keep the notation we developed for $\rep_{8,0}$.)
The representation $\rep_{9,1}$ is Weyl, since the volume element $\vol =
\ob_{-1} \vm \ob_0 \vm \cdots \vm \ob_8$ is represented by
\begin{equation}
\rep_{9,1}(\vol) = \sig3 \otimes \one = \pmatrix{ \one&\phm 0 \cr 0 &
-\one\cr}
= \rep_{-1} \rep_0 \ldots \rep_8 =: \rep_{11}.
\end{equation}
The Weyl projections (\ref{eq:Weyl projection}) take the form
\begin{equation}
\Pmau_+ = \pmatrix{\one & \zero\cr \zero & \zero\cr},
\qquad\Pmau_- = \pmatrix{\zero & \zero\cr \zero & \one\cr}.
\end{equation}
We denote an element $\carr \in \Carr = \O^4$ of the carrier space by its
Weyl projections
\begin{equation}
\carr_\pm := \Pmau_\pm \carr \in \O^2,
\end{equation}
where we discard the two vanishing components of $\carr_\pm$.
The identity
\begin{equation}
\Slash{x}\Slash{x} = x^\mu x_\mu\, \one \quad \iff \quad
\pmatrix{{\bf X}\tilde{\bf X} &\zero \cr  \zero &\tilde{\bf X}{\bf X}\cr}
= x^\mu x_\mu \pmatrix{ \one&\zero \cr\zero  &\one \cr},
\label{eq:rep property}
\end{equation}
holds under left multiplication because of the alternative property
(\ref{eq:associative inverse}), since only one full octonion $x$ and its
conjugate are contained in ${\bf X}$ and $\tilde{\bf X}$.
Noting that
\begin{equation}
\tilde{\bf X} = {\bf X} - (\tr{\bf X})\,\one,
\label{eq:cl tilde}
\end{equation}
it follows that
\begin{equation}
{\bf X}\tilde {\bf X} = {\bf X}^2 - (\tr{\bf X})\,{\bf X}
= \tilde {\bf X} {\bf X} = -\det{\bf X}\,\one = x^\mu x_\mu \,\one,
\label{eq:det}
\end{equation}
since the characteristic polynomial for a hermitian $2\times 2$-matrix $A$ is
$p_A(\lambda) = \lambda^2 - \tr{A} \lambda + \det A$.
Polarizing (\ref{eq:det}), we get
\begin{equation}\acsz\begin{array}{rl}
2 x_\mu y^\mu \,\one ={}& {\bf X}\tilde {\bf Y} + {\bf Y}\tilde {\bf X}
=\tilde {\bf X} {\bf Y} + \tilde {\bf Y} {\bf X}
\\ \noalign{\smallskip}
\iff\quad
2 g_{\mu\nu} \,\one ={}& \Rep_\mu\tilde\Rep_\nu + \Rep_\nu\tilde\Rep_\mu
=\tilde\Rep_\mu \Rep_\nu + \tilde\Rep_\nu \Rep_\mu\,.
\end{array}\end{equation}
To extract components, we have the familiar formulas involving traces:
\begin{equation}
x_\mu= {1\over 4}\Re\tr{\Slash{x} \rep_\mu}
= {1\over 4}\Re\tr{{\bf X}\tilde\Rep_\mu + \tilde {\bf X}\Rep_\mu}
= {1\over 2}\Re\tr{{\bf X}\tilde\Rep_\mu}
= {1\over 2}\Re\tr{\tilde {\bf X}\Rep_\mu}.\label{eq:invert rep}
\end{equation}
Considering
\begin{equation}
\rep_\mu = \left\{
\matrix{\phm \rep_\mu, & (\mu\not= -1)\cr
-\rep_\mu,& (\mu = -1)\cr}\right.\,,
\end{equation}
a hermitian form $\Sphf$ is given by
\begin{equation}
\Sphf(\carr) := \carr^\dagger \Sphf = \carr^\dagger \rep_{11} \rep_{-1}
= \matrix{ \pmatrix{ \carr_+^\dagger & \carr_-^\dagger \cr}\cr
\phantom{\carr_+^\dagger}\cr}
\pmatrix{\zero &\one \cr \one &\zero \cr}
= \pmatrix{ \carr_-^\dagger& \carr_+^\dagger\cr} =: \adj\carr\,.
\end{equation}
So the scalar covariant formed out of $\carr,\acarr \in \Carr$ is
\begin{equation}
\Sphf(\carr,\acarr) = \Re \adj\carr \acarr
= \Re \pmatrix{ \carr_-^\dagger& \carr_+^\dagger\cr}
\pmatrix{ \acarr_+ \cr \acarr_- \cr}
= \Re (\carr_-^\dagger\acarr_+ + \carr_+^\dagger\acarr_-),
\end{equation}
which only involves terms combining spinors of opposite chirality.
For the vector covariant $y$, we obtain
\begin{equation}\acsz\begin{array}{rl}
y_\mu :={}& \Re \adj\carr \rep_\mu\acarr
\\ \noalign{\medskip}
={}&\Re \left[
\matrix{\pmatrix{ \carr_-^\dagger & \carr_+^\dagger\cr}\cr
\phantom{\carr_+^\dagger}\cr}
\pmatrix{\zero & \Rep_\mu\cr \tilde\Rep_\mu & \zero\cr}
\pmatrix{ \acarr_+ \cr \acarr_- \cr}\right]
\\ \noalign{\medskip}
={}& \Re (\carr_+^\dagger \tilde\Rep_\mu \acarr_+
+ \carr_-^\dagger \Rep_\mu \acarr_-)
= \Re \tr{\acarr_+\carr_+^\dagger \tilde\Rep_\mu
+ \acarr_-\carr_-^\dagger \Rep_\mu }
\\ \noalign{\medskip}
={}& {1\over 2} \left[
\Re \tr{\acarr_+\carr_+^\dagger \tilde\Rep_\mu
+ \acarr_-\carr_-^\dagger \Rep_\mu }
+\Re \tr{(\acarr_+\carr_+^\dagger \tilde\Rep_\mu
+ \acarr_-\carr_-^\dagger \Rep_\mu)^\dagger}\right]
\\ \noalign{\medskip}
={}& {1\over 2}
\Re \tr{[\acarr_+\carr_+^\dagger +\carr_+\acarr_+^\dagger]\tilde\Rep_\mu
+ [\acarr_-\carr_-^\dagger + \carr_-\acarr_-^\dagger] \Rep_\mu }
\\ \noalign{\medskip}
={}& {1\over 2}
\Re \tr{[\acarr_+\carr_+^\dagger +\carr_+\acarr_+^\dagger]\tilde\Rep_\mu
+ \widetilde{[\acarr_-\carr_-^\dagger + \carr_-\acarr_-^\dagger]}
\tilde\Rep_\mu }\,.
\end{array}\end{equation}
So the vector covariant is formed of combinations of spinors of the same
chirality.  Since the hermitian matrix ${\bf Y}$ is completely determined by
the components according to (\ref{eq:invert rep}) and the terms in square
brackets are hermitian, we can give a formula analogous to (\ref{eq:vector
covariant}):
\begin{equation}\acsz\begin{array}{rl}
\Slash{y} ={}& \pmatrix{0&{\bf Y}\cr\tilde{\bf Y}&0\cr}
:= \rep^k \Re \adj\carr \rep_k\acarr
\\ \noalign{\smallskip}
={}& \pmatrix{\zero& [\acarr_+\carr_+^\dagger +\carr_+\acarr_+^\dagger]
+\widetilde{[\acarr_-\carr_-^\dagger + \carr_-\acarr_-^\dagger]}\cr
\widetilde{[\acarr_+\carr_+^\dagger +\carr_+\acarr_+^\dagger]}
+[\acarr_-\carr_-^\dagger + \carr_-\acarr_-^\dagger]&\zero\cr}\,.
\end{array}\end{equation}
Proper Lorentz transformations are generated by pairs of timelike (resp.\
spacelike) unit vectors $u,v \in V$, i.e., $u_\mu u^\mu = \mp 1 = v_\mu
v^\mu$.  We choose $v = \ob_{-1}$ fixed
\begin{equation}\acsz\begin{array}{rl}
\Sslash{x^\prime}{4} ={}& \Slash{u}\rep_{-1}\Slash{x}\rep_{-1}\Slash{u}
= \pmatrix{\zero & {\bf UXU}\cr \tilde{\bf U}\tilde{\bf X}\tilde{\bf U} &
\cr},
\\ \noalign{\smallskip}
\carr^\prime ={}& \Slash{u}\rep_{-1}\carr
= \pmatrix{ -{\bf U}\carr_+ \cr  \tilde{\bf U}\carr_- \cr}.
\end{array}\label{eq:SO(9,1) vectors and spinors}\end{equation}
The correct transformation behavior of spinors and vectors is ensured by the
Moufang identities as in the 8-dimensional case, since $\Slash{u}$ contains
additional real parameters but only one full octonion.  This form of proper
Lorentz transformations makes the isomorphism $SL(2,\O) \isom SO(9,1)$ as Lie
groups precise.

Since for $\Cc := \rep_{-1}\rep_0\rep_8 = -\sig2\otimes\sig2$
\begin{equation}
\Cc \rep_\mu = \Bar{\rep_\mu} \Cc \qquad (-1\leq \mu \leq 8),
\end{equation}
a ``charge conjugation'' operation is given by
\begin{equation}
\carr_\Cc := \Bar{\Cc(\carr)} = -\sig2\otimes\sig2\, \Bar\carr
= \pmatrix{ \phm\sig2\,\Bar{\carr_-} \cr -\sig2\,\Bar{\carr_+} \cr}\,,
\end{equation}
which must involve the opposite octonionic algebra as it was pointed in
(\ref{eq:Cc spinor}) and (\ref{eq:Cc problem}).  This transition to the
opposite algebra for spinors with opposite chirality may be useful in theories
with $N>1$ supersymmetry.

Of course, we may iterate the process of shrinking and extending of a
representation with $\rep_{9,1}$ as a starting point.  We can shrink it to
obtain representations of $\Cl_0(9,1) \isom \Cl(9,0) \isom \Cl(1,8)$ and from
there to $\Cl_0(9,0) \isom \Cl(0,8)$ and $\Cl_0(1,8) \isom \Cl(1,7)$.  Also an
extension to a representation of $\Cl(10,2)$ is possible.

\section{An octonionic description of the Chevalley algebra and triality}
\label{sec:triality}

The triality automorphisms of the Chevalley algebra are well known and have
been discussed in detail before \cite{Cl:Chevalley,Cl:Adams,Cl:Porteous}, even
in an octonionic formulation \cite{Cl:Dundarer triality}.  However, in our
opinion, the following treatment based on the preparatory work of section
\ref{sec:8} adds another unique and very transparent perspective
with regard to this topic.

In the case of 8 euclidean dimensions we are in a special situation; the
spaces of vectors, $V$, even spinors, $\Spsp_0$, and odd spinors, $\Spsp_1$,
have the same dimension, namely 8.  This allows the construction of the
triality maps that interchange the transformation behavior of these three
spaces.  We define the Chevalley algebra $\Calg := V \oplus \Spsp_0 \oplus
\Spsp_1$ to be the direct sum of these three spaces.  This definition
automatically provides a vector space structure for $\Calg$.
Furthermore, $\Calg$ inherits an $SO(8)$-invariant bilinear form
$\Cbl = 2\,g \oplus 2\,\Sphf$ from the metric $g$ on the vector space and the
hermitian form $\Sphf$ on $\Spsp = \Spsp_0 \oplus \Spsp_1$.  (For notational
convenience later on, we put in a factor of 2 in the definition of $\Cbl$.)
For $a = a_v \oplus a_0 \oplus a_1$, $b = b_v \oplus b_0 \oplus b_1 \in
\Calg$, we obtain
\begin{equation}
\Cbl(a,b) = 2\,g(a_v,b_v) +
2\,\Sphf(\pmatrix{ a_0\cr \Bar{a_1}  \cr},\pmatrix{ b_0\cr \Bar{b_1}  \cr})
= 2\,\Re( a_v \Bar{b_v}  + \Bar{a_0} b_0 + a_1 \Bar{b_1}),
\label{eq:confirms}
\end{equation}
where we used the parametrization of the spinor components introduced in
section~\ref{subsec:twisted spinors}.  (\ref{eq:confirms}) confirms that
$\Sphf$ decomposes and is a real symmetric bilinear form on the 16 real spinor
components.  The $SO(8)$-invariance of $\Cbl$ is clear using the results of
section~\ref{subsec:8D trafos}.  Furthermore, we observed in (\ref{eq:scalar})
that the expression
\begin{equation}
\Ctl^\prime(a) := \Re \adj{a_1} \Sslash{\Bar{a_v}}{6} a_0
= \Re \left[
\matrix{\pmatrix{ 0 & a_1\cr}\cr
\phantom{a_1}\cr}
\pmatrix{0 &\Bar{a_v}\cr a_v& 0\cr}
\pmatrix{ a_0 \cr 0\cr}\right]
= \Re a_1a_va_0
\end{equation}
is $SO(8)$-invariant.  (Note that, we also redefined our basis of $V$ by
octonionic conjugation for symmetry reasons, which will become relevant
below.)  By polarization, we define a $SO(8)$-invariant symmetric trilinear
form on $\Calg$, which we denote by $\Ctl$:
\begin{equation}
\Ctl(a,b,c) :=
\Re (a_1b_vc_0+a_1c_vb_0+b_1a_vc_0+b_1c_va_0+c_1a_vb_0+c_1b_va_0)
\;\quad (a,b,c \in \Calg).
\end{equation}
The Chevalley product ``$\circ_\Calg$'' is then implicitly defined to satisfy
the following condition connecting $\Cbl$ and $\Ctl$:
\begin{equation}
\Cbl(a \circ_\Calg b,c) = \Ctl(a,b,c) \qquad \forall\,a,b,c \in \Calg.
\label{eq:Calg product}
\end{equation}
The Chevalley product is obviously symmetric and $SO(8)$ invariant.

In this setting the triality maps are just automorphisms of the Chevalley
algebra, which interchange $V$, $\Spsp_0$, and $\Spsp_1$.
But before we describe the triality maps, we will take advantage of the
octonionic formalism and rewrite the bilinear and trilinear forms, $\Cbl$ and
$\Ctl$, and the Chevalley product by representing elements of the Chevalley
algebra by octonionic hermitian $3\times 3$-matrices with vanishing diagonal
elements,
\begin{equation}
a = \pmatrix{0&\Bar{a_v}&a_0\cr a_v&0&\Bar{a_1}\cr \Bar{a_0}&a_1&0\cr}
= \pmatrix{\Sslash{\Bar{a_v}}{6} & a_\sp\cr \adj{a_\sp} & 0\cr}
\in \Calg,
\end{equation}
where $a_\sp = \left({a_0\atop\Bar{a_1}}\right) = a_o \oplus a_1 \in \Spsp$.
Then the bilinear form $\Cbl$ is given by
\begin{equation}\acsz\begin{array}{rl}
\Cbl(a,b) ={}& {1\over 2}\tr{ab+ba} = \tr{a \circ b}
\\ \noalign{\smallskip}
={}& {1\over 2}\tr{
\pmatrix{0&\Bar{a_v}&a_0\cr a_v&0&\Bar{a_1}\cr \Bar{a_0}&a_1&0\cr}
\pmatrix{0&\Bar{b_v}&b_0\cr b_v&0&\Bar{b_1}\cr \Bar{b_0}&b_1&0\cr}
+\pmatrix{0&\Bar{b_v}&b_0\cr b_v&0&\Bar{b_1}\cr \Bar{b_0}&b_1&0\cr}
\pmatrix{0&\Bar{a_v}&a_0\cr a_v&0&\Bar{a_1}\cr \Bar{a_0}&a_1&0\cr}}
\\ \noalign{\medskip}
={}& {1\over 2}\tr{
\pmatrix{\Bar{a_v}b_v+a_0\Bar{b_0}&a_0b_1&\Bar{a_v}\Bar{b_1}\cr
\Bar{a_1}\Bar{b_0}&a_v\Bar{b_v}+\Bar{a_1}b_1&a_vb_0\cr
a_1b_v&\Bar{a_0}\Bar{b_v}&\Bar{a_0}b_0+a_1\Bar{b_1}\cr}}
\\ \noalign{\medskip}
&\qquad
{}+{1\over
2}\tr{\pmatrix{\Bar{b_v}a_v+b_0\Bar{a_0}&b_0a_1&\Bar{b_v}\Bar{a_1}\cr
\Bar{b_1}\Bar{a_0}&b_v\Bar{a_v}+\Bar{b_1}a_1&b_va_0\cr
b_1a_v&\Bar{b_0}\Bar{a_v}&\Bar{b_0}a_0+b_1\Bar{a_1}\cr}}
\\ \noalign{\medskip}
={}&{1\over 2}[(\Bar{a_v}b_v+\Bar{b_v}a_v+a_v\Bar{b_v}+b_v\Bar{a_v})
+(a_0\Bar{b_0}+b_0\Bar{a_0}+\Bar{a_0}b_0+\Bar{b_0}a_0)
\\ \noalign{\smallskip}
&\phantom{{1\over 2}[}
{}+( \Bar{a_1}b_1+\Bar{b_1}a_1+a_1\Bar{b_1}+b_1\Bar{a_1})]
\\ \noalign{\smallskip}
={}&2\,\Re( a_v \Bar{b_v}  + \Bar{a_0} b_0 + a_1 \Bar{b_1}),
\end{array}\label{eq:Cbilinear}\end{equation}
where ``$\circ$'' denotes the symmetrized matrix product
\begin{equation}
a \circ b := {1\over 2}(ab+ba).
\end{equation}
In fact, the symmetrized product is the Jordan product and the matrices that
we are dealing with are a subset of the exceptional Jordan algebra of $3\times
3$ octonionic hermitian matrices \cite{Cl:Jordan}.

For the trilinear form $\Ctl$ we find
\begin{equation}\acsz\begin{array}{rl}
\Ctl(a,b,c) ={}& \tr{(a\circ b)\circ c}
\\ \noalign{\smallskip}
={}& {1\over 4}[
(a_0b_1c_v+c_va_0b_1+\Bar{b_1}\Bar{a_0}\Bar{c_v}+\Bar{c_v}\Bar{b_1}\Bar{a_0})+
(b_0a_1c_v+c_vb_0a_1
\\ \noalign{\smallskip}&{}
+\Bar{a_1}\Bar{b_0}\Bar{c_v}+\Bar{c_v}\Bar{a_1}\Bar{b_0})+
(a_vb_0c_1+c_1a_vb_0+\Bar{b_0}\Bar{a_v}\Bar{c_1}+\Bar{c_1}\Bar{b_0}\Bar{a_v})
\\ \noalign{\smallskip}&{}
+(b_va_0c_1+c_1b_va_0+\Bar{a_0}\Bar{b_v}\Bar{c_1}+\Bar{c_1}\Bar{a_0}\Bar{b_v})
+(a_1b_vc_0+c_0a_1b_v
\\ \noalign{\smallskip}&{}
+\Bar{b_v}\Bar{a_1}\Bar{c_0}+\Bar{c_0}\Bar{b_v}\Bar{a_1})
+(b_1a_vc_0+c_0b_1a_v+\Bar{a_v}\Bar{b_1}\Bar{c_0}+\Bar{c_0}\Bar{a_v}\Bar{b_1})]
\\ \noalign{\smallskip}
={}&\Re (b_1c_va_0+a_1c_vb_0+c_1a_vb_0+c_1b_va_0+a_1b_vc_0+b_1a_vc_0).
\end{array}\label{eq:Ctrilinear}\end{equation}
It follows from (\ref{eq:Calg product}), (\ref{eq:Cbilinear}), and
(\ref{eq:Ctrilinear}) that the Chevalley product ``$\circ_\Calg$'' is given by
the off-diagonal elements of the symmetrized matrix product ``$\circ$'',
\begin{eqnarray}
&\tr{(a \circ_\Calg b) \circ c} = \Cbl(a \circ_\Calg b,c)
= \Ctl(a,b,c) = \tr{(a \circ b) \circ c}
\\ \noalign{\smallskip}\label{eq:derive CJproduct}
\implies &
(a \circ_\Calg b) = (a \circ b)_\Calg,
\end{eqnarray}
where the subscript ``$\Calg$'' on a matrix denotes the matrix with erased
diagonal elements, i.e.,
\begin{equation}
(a \circ b)_\Calg :=
\pmatrix{0&a_0b_1+b_0a_1&\Bar{a_v}\Bar{b_1}+\Bar{b_v}\Bar{a_1}\cr
\Bar{a_1}\Bar{b_0}+\Bar{b_1}\Bar{a_0}&0&a_vb_0+b_va_0\cr
a_1b_v+b_1a_v&\Bar{a_0}\Bar{b_v}+\Bar{b_0}\Bar{a_v}&0\cr}\,.
\label{eq:CJproduct}
\end{equation}
(Note that only the off diagonal elements of $a \circ b$ contribute to the
last term of (\ref{eq:derive CJproduct})).
Traditionally the Chevalley product is written in terms of Clifford products,
which we combine into the $3\times 3$-matrix
\begin{equation}
a \circ_\Calg b =
\pmatrix{\Rep^k \adj{a_\sp}\Rep_k b_\sp
& \Sslash{\Bar{a_v}}{6}b_\sp + \Sslash{\Bar{b_v}}{6}a_\sp\cr
\adj{a_\sp}\Sslash{\Bar{b_v}}{6} +\adj{b_\sp}\Sslash{\Bar{a_v}}{6}&0\cr}
\end{equation}
What we have done is to utilize the Jordan product and project onto the
Chevalley algebra.  Since both $\Cbl$ and $\Ctl$ are expressed entirely in
terms of the Jordan product, automorphisms of the Jordan product, that map the
Chevalley algebra onto itself, will also be automorphisms of the Chevalley
algebra.  We have already encountered one such automorphism, namely the
orthogonal transformation corresponding to a generator $\fp_v \in V$ with
$\normsq{\fp_v} = 1$, which is written in matrix form
\begin{equation}
\tau_{\fp_v}(a) :=
\pmatrix{ 0&\Bar{\fp_v}&0\cr \fp_v&0&0\cr 0&0&1\cr}
\pmatrix{0&\Bar{a_v}&a_0\cr a_v&0&\Bar{a_1}\cr \Bar{a_0}&a_1&0\cr}
\pmatrix{ 0&\Bar{\fp_v}&0\cr \fp_v&0&0\cr 0&0&1\cr}
= \pmatrix{
\Sslash{\Bar{\fp_v}}{6}\Sslash{\Bar{a_v}}{6}\Sslash{\Bar{\fp_v}}{6}
& \Sslash{\Bar{\fp_v}}{6}a_\sp\cr
\adj{a_\sp}\Sslash{\Bar{\fp_v}}{6}& 0\cr}\,.
\label{eq:triality map v}
\end{equation}
This first triality map combines the vector action and spinor action of the
Clifford group (see section~\ref{subsec:Cg}).  The action of the generator
$\fp_v$ is a reflection at a hyperplane orthogonal to $\fp_v$ combined with an
inversion of the whole space.  This transformation is an improper rotation and
interchanges even and odd spinors:
\begin{equation}\acsz\begin{array}{rll}
\tau_{\fp_v}(a_v) ={}& \fp_v\Bar{a_v} \fp_v \quad&\in V\,,
\\ \noalign{\smallskip}
\tau_{\fp_v}(a_0) ={}& \Bar{(\fp_v a_0)} &\in \Spsp_1\,,
\\ \noalign{\smallskip}
\tau_{\fp_v}(a_1) ={}& \Bar{(a_1 \fp_v)} &\in \Spsp_0\,.
\end{array}\end{equation}
Using the Moufang identities, it is easy to check that $\tau_{\fp_v}$ is
indeed an automorphism of $\Calg$ of order 2, i.e., $\tau_{\fp_v}^2 = 1$.
Composing an even number of maps $\tau_{\fp_v}$ with different parameters
$\fp_v$, we generate the simple orthogonal group $SO(8)$ as is seen in
(\ref{eq:vector SO(8)}) and (\ref{eq:spinor SO(8)}).
{}From the form of (\ref{eq:triality map v}), it is obvious that there are two
more families of automorphisms of $\Calg$ of order 2, parametrized by an even
spinor variable $\fp_0$ and an odd spinor variable $\fp_1$ with
$\normsq{\fp_0} = 1 = \normsq{\fp_1}$:
\begin{equation}
\tau_{\fp_0}(a) :=
\pmatrix{ 0&0&\fp_0\cr 0&1&0\cr \Bar{\fp_0}&0&0\cr}
\pmatrix{0&\Bar{a_v}&a_0\cr a_v&0&\Bar{a_1}\cr \Bar{a_0}&a_1&0\cr}
\pmatrix{ 0&0&\fp_0\cr 0&1&0\cr \Bar{\fp_0}&0&0\cr}
\label{eq:triality map 0}
\end{equation}
and
\begin{equation}
\tau_{\fp_1}(a) :=
\pmatrix{ 1&0&0\cr 0&0&\Bar{\fp_1}\cr 0&\fp_1&0\cr}
\pmatrix{0&\Bar{a_v}&a_0\cr a_v&0&\Bar{a_1}\cr \Bar{a_0}&a_1&0\cr}
\pmatrix{ 1&0&0\cr 0&0&\Bar{\fp_1}\cr 0&\fp_1&0\cr}\,.
\label{eq:triality map 1}
\end{equation}
For these two families of maps, the matrix formalism shows the clear parallel
structure to the maps $\tau_{\fp_v}$.
Traditionally expressions in terms of both Clifford products and the spinor
bilinear form are used for the maps $\tau_{\fp_0}$ and $\tau_{\fp_1}$, which
obscures this symmetry, because in $\tau_{\fp_v}$ only Clifford products are
used.  These two families preserve one of the spinor spaces and interchange
the other one with $V$:
\begin{equation}\acsz\begin{array}{rll}
\tau_{\fp_0}(a_v) ={}& \Bar{(a_v \fp_0)}&\in \Spsp_1\,,
\\ \noalign{\smallskip}
\tau_{\fp_0}(a_0) ={}& \fp_0\Bar{a_0} \fp_0 \quad&\in \Spsp_0\,,
\\ \noalign{\smallskip}
\tau_{\fp_0}(a_1) ={}& \Bar{(\fp_0a_1)} &\in V\,,
\end{array}\end{equation}
and
\begin{equation}\acsz\begin{array}{rll}
\tau_{\fp_1}(a_v) ={}& \Bar{(\fp_1a_v)}&\in \Spsp_0\,,
\\ \noalign{\smallskip}
\tau_{\fp_1}(a_0) ={}& \Bar{(a_0 \fp_1)} &\in V\,,
\\ \noalign{\smallskip}
\tau_{\fp_1}(a_1) ={}& \fp_1\Bar{a_1}\fp_1\quad &\in \Spsp_1\,,
\end{array}\end{equation}
By combining two triality maps with the same octonionic parameter $\fp_v = \fp
=
\fp_0$ from different families, we obtain a automorphisms $\Xi_\fp$ of
order~3:
\begin{equation}\acsz\begin{array}{rc}
&\Xi_\fp = \tau_{\fp_v=\fp} \circ \tau_{\fp_0=\fp}
\\ \noalign{\smallskip}
\implies \quad&
\Xi_\fp(a) = \pmatrix{ 0&1&0\cr 0&0&\Bar{\fp}\cr \fp&0&0\cr}
\pmatrix{0&\Bar{a_v}&a_0\cr a_v&0&\Bar{a_1}\cr \Bar{a_0}&a_1&0\cr}
\pmatrix{ 0&0&\Bar{\fp}\cr 1&0&0\cr 0&\fp&0\cr}\qquad(a \in \Calg),
\end{array}\end{equation}
hence
\begin{equation}\acsz\begin{array}{rll}
\Xi_\fp(a_v) ={}& \Bar{\fp}a_v&\in \Spsp_0\,,
\\ \noalign{\smallskip}
\Xi_\fp(a_0) ={}& \fp a_0\fp &\in \Spsp_1\,,
\\ \noalign{\smallskip}
\Xi_\fp(a_1) ={}& \Bar{\fp}a_1\quad &\in V\,.
\end{array}\end{equation}
As is seen from their matrix forms, $\tau_{\fp_v=\fp}$ and $\Xi_\fp$ generate
$\perm_3$, the permutation group on three letters.  (In particular for $\fp=1$,
this is easy to verify.)  We observed before that the maps $\tau_{\fp_v}$
generate $O(8)$, so that the triality maps, we have found so far, have a group
structure isomorphic to $\perm_3\times SO(8)$.  It is known (see
\cite{Cl:Adams}) that this is the full automorphism group of the Chevalley
algebra, which is also the automorphism group of $SO(8)$.
This concludes our demonstration of triality.

\section{Finite vs.\ infinitesimal generators}
\label{sec:full group}

In this article we characterize orthogonal groups in terms of a set of finite
generators.  This approach is not as widely used as the description in terms
of infinitesimal generators, i.e., the Lie algebra of the group.  In this
section we compare the two approaches.

If we want to compare two Lie groups given by infinitesimal generators we know
how to proceed \cite{Cl:Wybourne}.  We determine their Lie algebra by working
out the commutators of the generators.  We then determine their structure
constants and identify the Lie algebra.  For semi-simple Lie algebras the
Cartan-Weyl normalization provides a unique identification.  We may also use a
Lie algebra homomorphism and determine its image and kernel to relate the two
groups in question.  Whether the homomorphism is surjective and injective can
often be determined by counting the dimension of the Lie algebras involved.
Having identified the Lie algebra we have full knowledge of the local
structure of the Lie group.  From this information we can construct the simply
connected universal covering group, which has this local structure.  However,
the Lie group we are trying to characterize may be neither connected nor
simply connected.  So in order to compare two groups we need to have some
global information about them in addition to the infinitesimal generators.

In section~\ref{subsec:Cg} we compared two groups given by finite generators,
namely the orthogonal group generated by reflections on hyperplanes and the
Clifford group generated by non-null vectors of the Clifford algebra.  The
relationship was established considering a group homomorphism.  The
homomorphism is surjective if the generators lie in the image.  This is the
analogue to counting the dimension of the Lie algebras.  Determining the
kernel, which has to be a normal subgroup, completes the comparison.  The
advantage of finite generators is the global information that they carry.
Having found an isomorphism based on the finite generators, we know that the
groups have the same global structure.

Even though the two descriptions have different features, they are closely
related.  The exponential map provides a means to parametrize a neighborhood
of the identity element of the group.  This coordinate chart can be translated
by a finite element in this neighborhood, hence we can construct an atlas of
the component of the group that is connected to the identity.  Actually, we
need information about the global structure to patch the charts together
correctly.  For an additional component of the group that is not connected to
the identity, we may use the same atlas, since the components are
diffeomorphic.

The finite generators that determine the groups considered in this article are
elements of a topological manifold of dimension less than the dimension of the
group.  For example, the octonions that generate $SO(8)$ (\ref{eq:vector
SO(8)}) are elements of the octonionic unit sphere, $S^7$.  Translating a disk
centered at a point $p \in S^7$ by $p^{-1} \in S^7$, we obtain a submanifold
of the group containing the identity.  (A generating set of a group is always
assumed to contain inverses of every element.)  This submanifold is of lower
dimension than the Lie group, so its tangent space at the identity is only a
linear subspace of the Lie algebra.  In most of our examples it is sufficient
to consider the translation of a sufficient number of disks contained in the
generating set to obtain linear subspaces that span the Lie algebra.
Otherwise the process continues by taking products of elements of two disks
around $p_1$ and $p_2$ in the generating set and translating these products by
$(p_1p_2)^{-1}$ to the identity.  An example of this latter construction is
the $S^6$ generating $SO(8)$ described in \cite{Cl:Lorentz}.  In this way
infinitesimal generators can be found starting from finite ones.

There is also a formal construction of the entire group; namely, the group is
given by the set of equivalence classes of finite sequences of generators.
The group product of two elements $[g_1], [g_2]$ is just the class of the
juxtaposition $[g_1g_2]$ of two representatives.  For the octonionic
description we need to do this decomposition into generators to find spinor
and vector transformations that are consistent.  For example, if a vector
given by $x \in \O$ transforms by $x \mapsto ux\Bar{u}$, which is an $SO(8)$
transformation, we need to re-express $ux\Bar{u}$ as $v_1(v_2(\ldots(v_k x
v_k)\ldots)v_2)v_1$ with $\normsq{v_1} = \normsq{v_2} = \cdots = \normsq{v_k}$
in order to determine the corresponding spinor transformation $\carr \mapsto
v_1(v_2(\ldots(v_k \carr)\ldots))$.
In general, octonionic transformations, because of their non-associativity,
involve this nesting of multiplications. Therefore the octonionic description
of Lie groups in terms of generators is the natural one.  Octonionic
descriptions of Lie algebras, which are also possible, have the disadvantage
that the exponential map no longer works because of the non-associativity.  So
this avenue does not provide a construction of finite group elements.

\section{Conclusion}\label{sec:conclusion}

We have demonstrated that the abstract octonionic algebra is a suitable
structure to represent Clifford algebras in certain dimensions.  We obtained
most of our results from the basic property of composition algebras, which is
the norm compatibility of multiplication, and its consequence alternativity.
The alternative property, in particular in the form of the Moufang identities,
was found to be responsible for ensuring the correct transformation behavior of
octonionic spinors and for ensuring the consistency of the representation in
terms of left multiplication by octonionic matrices.  The choice of a
multiplication rule for the octonions, in particular, the modified
``$X$-product'', was found to be related to coordinate transformations or a
change of basis of the spinor space.  The opposite octonionic algebra was
shown to be connected to an analogue of the charge conjugate representation.
The Clifford group and its action on vectors and spinors led to
octonionic representations of orthogonal groups in corresponding dimensions.
The natural octonionic description of these groups is in terms of generating
sets of the Lie group rather than in terms of generators of the Lie algebra.
This is due to the nested structure which is necessary to accommodate the
non-associativity of the octonions.

The usefulness of this tool of octonionic representations was evident in the
presentation of the triality automorphisms of the Chevalley algebra.  This
presentation unequivocally showed that the spaces of vectors and even and odd
spinors are interchangeable in this case.  We expect that a similar, fully
octonionic treatment of supersymmetrical theories will make their symmetries
more transparent.  In fact, we have successfully applied the methods of this
article to the CBS-superparticle \cite{Cl:superparticle}.  We hope to be able
to find a parallel treatment of the Green-Schwarz superstring.

\acknowledgments

JS is grateful to Dr.~Burt Fein for his comments and for pointing out
references concerning group theory and algebra.
JS wishes to thank the Bayley family for establishing the Bayley Graduate
Fellowship, which he had the honor to receive this year.  This work was
partially funded by NSF grant \# PHY92-08494.

\end{document}